\documentclass[12pt,preprint]{aastex}

\begin{document}

\title{The VLA Low-frequency Sky Survey}

\author{A.~S.~Cohen \altaffilmark{1},
W.~M.~Lane \altaffilmark{1},
W.~D.~Cotton \altaffilmark{2},
N.~E.~Kassim \altaffilmark{1},
T.~J.~W.~Lazio \altaffilmark{1},
R.~A.~Perley \altaffilmark{3},
J.~J.~Condon \altaffilmark{2},
W.~C.~Erickson \altaffilmark{4},
}

\altaffiltext{1}{Naval Research Laboratory, Code 7213, Washington, DC, 
20375 USA, Aaron.Cohen@nrl.navy.mil, Wendy.Lane@nrl.navy.mil}
\altaffiltext{2}{National Radio Astronomy Observatory, 520 Edgemont Road, 
Charlottesville, VA, 22903 USA}
\altaffiltext{3}{National Radio Astronomy Observatory, P.O. Box 0, 
Socorro, NM 87801 USA}
\altaffiltext{4}{School of Mathematics and Physics, University of Tasmania, 
Hobart, TAS 7005, Australia}

\begin{abstract}
The Very Large Array (VLA) Low-frequency Sky Survey (VLSS) has imaged 
95\% of the $3\pi$ sr of sky north of $\delta = -30^\circ$ at a 
frequency of 74~MHz (4~meter wavelength). 
The resolution is 80$''$ (FWHM) throughout, and the typical RMS noise level 
is $\langle \sigma \rangle \approx 0.1$~Jy/beam.  The typical 
point-source detection limit is 0.7~Jy/beam and so far nearly 70,000 
sources have been catalogued.  This survey used the 74~MHz system added to 
the VLA in 1998.  It required new imaging algorithms to remove the large 
ionospheric distortions at this very low frequency throughout the entire
$\sim 11.9^\circ$ field of view.  This paper describes the observation 
and data reduction methods used for the VLSS and presents the survey 
images and source catalog.  All of the calibrated images and the source 
catalog are available online (URL={\tt http://lwa.nrl.navy.mil/VLSS}) 
for use by the astronomical community.

\end{abstract}


\section{Introduction}

Recently, increasingly powerful telescopes and data-reduction abilities have 
made it possible to complete comprehensive and sensitive radio surveys, 
notably the 325~MHz Westerbork Northern Sky Survey 
\citep[WENSS;][]{rengelink97}, 
the 843~MHz Sydney University Molonglo Sky Survey 
\citep[SUMSS;][]{bock99}, the 1.4~GHz NRAO VLA Sky Survey 
\citep[NVSS;][]{condon98} and the 1.4~GHz Faint Images of the Radio Sky at 
Twenty-cm survey \citep[FIRST;][]{becker95,white97}.  These four 
surveys have resulted in the detection of millions of radio sources.  Their
data have already made a valuable contribution to topics such as the nature of
extragalactic radio sources and their relation to galaxy formation, the 
large-scale structure of the universe, and the use of Galactic foreground 
polarization as a probe of the interstellar medium in our own Galaxy.  In 
addition, the simple web access to maps and source information of the NVSS 
and FIRST databases has made it easy for researchers with little or no 
experience observing at radio frequencies to use this information in their 
work.

Until recently, it was impossible to make images approaching the dynamic
range and angular resolution of WENSS, SUMSS, and the NVSS at frequencies 
below 150 MHz owing to the severe ionospheric phase changes at such low 
frequencies.  This changed with the development of a 74~MHz 
(4-meter wavelength) system on the VLA \citep{kassim93}, which 
enabled sub-arcminute resolution synthesis 
imaging with a connected-element interferometer below 150~MHz for the first 
time.  Fully implemented in 1998, this new system has produced interesting 
science and provided valuable experience in the challenges associated with 
low-frequency observing at high angular resolution.  For a complete 
description of the VLA 74~MHz system and its capabilities, see 
\citet{kassim07}.

Phase distortions from the ionosphere and a large field of view introduce 
a particularly difficult problem for high-resolution imaging at 74~MHz.  
The isoplanatic patch is defined as a region on the sky small enough that 
angular variations in the ionospheric phase distortions across it are 
negligible.  Unlike at higher frequencies, the isoplanatic patch at 74~MHz 
for the VLA A and B-configurations (with maximum baselines of 36 and 11~km 
respectively) is significantly smaller than the field of view.  Therefore, 
initially the only sources that could be imaged were those strong enough 
that all other sources in the field of view outside of its isoplanatic 
patch were weak enough in comparison to be ignored during calibration.  
This restricted the system to 
sources with flux densities of at least $\sim$100~Jy at 74~MHz.  This 
obstacle has recently been greatly reduced through the development of new 
calibration algorithms (described in Section~\ref{reduction.sec}) and the 
availability of the necessary computational power to implement them.  It is
now possible to image an entire field of view and detect sources as weak 
as $\sim$0.1~Jy during most circumstances for the 11~km B-configuration and 
in many cases for the 35~km A-configuration.

These new algorithms have greatly extended the scientific capability of the 
74~MHz VLA system, and have made it possible to conduct 
efficient surveys.  In 2002 we began the VLA Low-frequency Sky Survey (VLSS), 
a 74~MHz survey of the entire sky north of $\delta > -30^\circ$.  VLSS 
images have a resolution of $80''$ and a median map RMS noise level of 
$\sim$0.1~Jy/beam.  The specific survey parameters are summarized in 
Table~\ref{tab.parms}.  This paper describes the methods and presents the 
results (images and a source catalog) of the VLSS, which is now 95\% complete.
In a future paper we will present analysis of these results and their 
scientific implications.

\section{Survey Motivation}

There are several scientific motivations for a new, relatively high 
resolution, high sensitivity survey at 74~MHz.  The first is the study of 
ultra-steep-spectrum objects ($\alpha < -1.3$).  These include halos and 
relics in clusters of galaxies \citep{ensslin02}, fossil radio galaxies 
\citep{slee01}, high redshift radio galaxies \citep{chambers87,debreuck00} 
and pulsars.  The VLSS has the potential not only to help study known examples 
of such objects, but also to help in the discovery of new objects in these 
classes.  The second main advantage of a 74~MHz survey is that the spectra 
of known objects can be extended to a frequency low enough that extrinsic 
(eg. free-free absorption) and intrinsic (eg. synchrotron self-absorption, 
electron energy-spectral cutoffs, free-free absorption) spectral effects 
can be distinguished \citep{kassim95}.  This is useful for studying 
physical processes related to acceleration, turbulence, and propagation in 
normal galaxies, supernova remnants, HII regions, and the interstellar 
medium.  Third, extragalactic samples selected 
at 74~MHz are dominated by isotropic (lobe-dominated) radio 
emission, unlike those found at higher frequencies.  A 74~MHz sample 
thus provides an unbiased view of the parent populations used in 
``unification'' models to account for the diverse source populations 
observed at higher frequencies \citep[eg.][]{wall97}.  Finally, a large 
survey that pushes the phase space of 
previous observations is often useful in uncovering rare, but potentially 
important new phenomena.

On a technical level, the VLSS
will produce a low-frequency sky model and an initial calibration grid for 
future low-frequency telescopes, such as the Low Frequency Array (LOFAR) and
the Long Wavelength Array (LWA).  The VLSS catalog will also help in the 
design of future low-frequency telescopes since the still-unsolved problem
of calibrating at low frequencies on long baselines ($>$50~km) will likely 
depend on the expected number of calibrator sources available within a 
field of view.  

\section{Observation Strategy}

Observations were carried out in four observational programs over
the course of five years (see Table ~\ref{tab:Obsdates}).  They are
substantially complete at this time; 10 hours remain in each of the
upcoming B- and BnA-configurations to re-observe a few remaining
high-noise fields.  Overall we have used just over 900 hours of VLA
time for this survey.  During each phase of observations every effort
was made to observe contiguous sky areas so that the intermediate
catalog releases would be as useful as possible for science.

\subsection{Pointing Grid}

\begin{figure}
\plotone{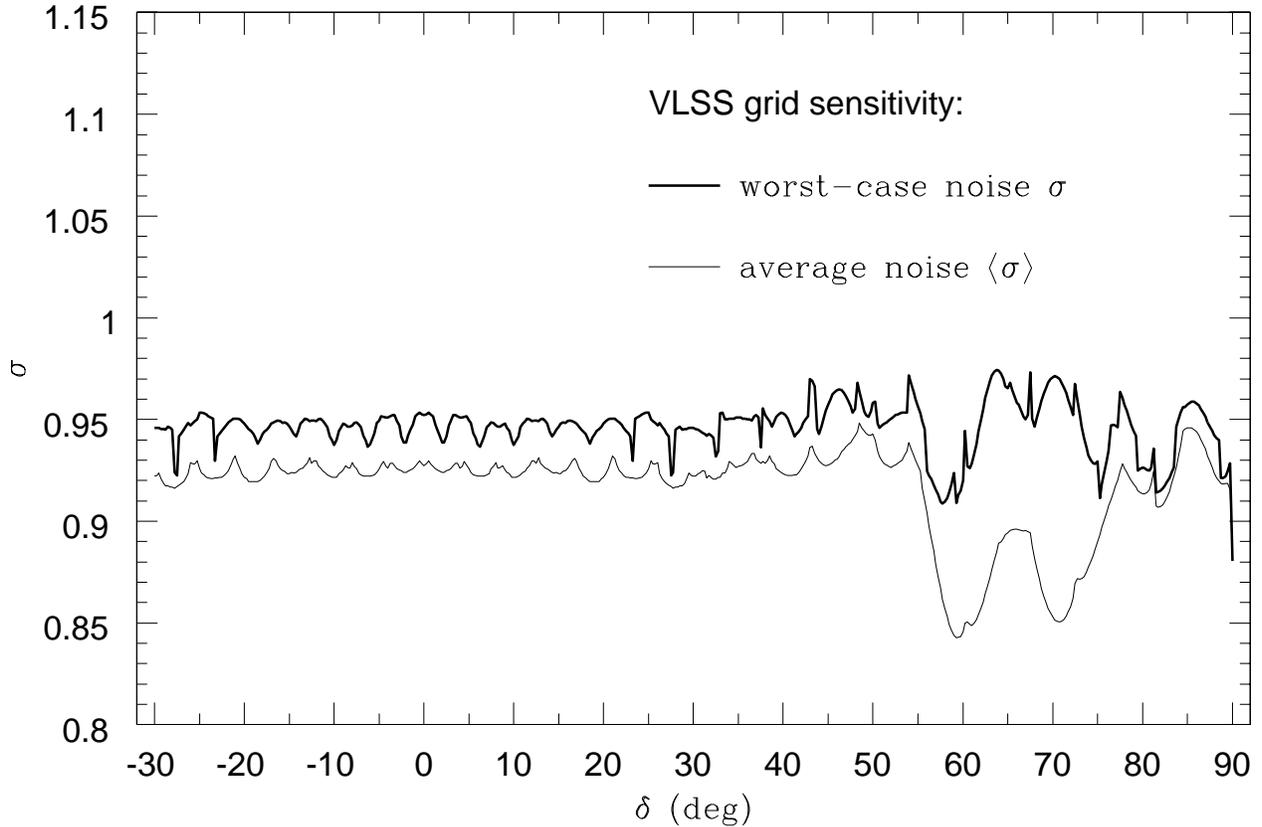}
\caption{A plot showing the average (light line) and worst
case (bold line) RMS noise (y-axis) as a function of
declination (x-axis), based on our VLSS pointing grid and the shape of the 
74 MHz VLA primary beam.  This assumes a constant RMS noise level for 
each field, with $\sigma = 1$ normalized to the noise level at the center 
of any individual pointing.  Due to the hexagonal observing pattern, the 
effects shown are not a simple function of declination.
The worst case noise is only slightly higher than the average,
indicating that the VLSS pointing grid produces essentially uniform 
sensitivity across the entire region surveyed.  
\label{grid.weights.fig}}
\end{figure}

The smallest element of the VLSS is the image of a single field of view 
surrounding a given pointing center.  The primary-beam sensitivity pattern 
of a 25-meter VLA dish at 74~MHz is a circular Gaussian with a 
full-width at half maximum (FWHM) diameter of 
$11.9^{\circ}$. 
In order to obtain uniform sensitivity over the entire survey area,
overlapping observations were made on a roughly hexagonal
grid of 523 pointing centers covering the entire sky north of 
$\delta > -30^{\circ}$.  The grid spacing was
\begin{equation}
\Delta \approx \theta_p/\sqrt{2} \approx 8.6^{\circ}, 
\end{equation}
where $\theta_p$ is the FWHM primary beamwidth.  
The partially overlapping images of each field are
weighted to correct for the primary-beam attenuation and combined
to produce the final sky images as described by \citet{condon98}.
For the same RMS noise level at the center of each pointing,
this grid produces nearly uniform sensitivity 
(Figure~\ref{grid.weights.fig}).  

\subsection{Resolution}
Fields in the declination range $-10^{\circ} < \delta < 80^{\circ}$
were observed in the B-configuration which produces a dirty-beam size
of between $60''$ and $80''$ at this frequency.  In order to produce 
uniform resolution and avoid highly elongated beams for fields that 
never reach high elevations, all
fields at declinations $\delta > 80^{\circ}$ and $\delta < -10^{\circ}$ 
were observed in the BnA-configuration which provides much longer 
baselines along the north-south axis than along the east-west axis.
All images were restored with a circular $80''$ FWHM beam for uniform 
resolution throughout the entire survey region.

\subsection{74 MHz Specific Considerations} 
The observations had a bandwidth of 1.56~MHz centered on the radio 
astronomy allocation of 73.8~MHz.  In order to facilitate the excision of
radio-frequency interference (RFI) and minimize bandwidth smearing, we 
observed in multi-channel continuum mode with 128 channels after online 
Hanning smoothing.  We used an integration time of 10 seconds, the smallest 
available on this system, which tangentially convolves the point-source 
response by a function whose width is proportional to the distance from the 
phase center and reaches 15$''$ at the primary half-power circle.  Convolved
with the 80$''$ restoring beam, this yields a point-source response that is 
extended to 81.4$''$ in the tangential direction at the primary half-power 
point.

Ionospheric distortions increase with decreased elevation.  Therefore
we required that all fields in the B-configuration be observed at 
elevations  $\geq30^{\circ}$, and the lower-declination 
BnA-configuration fields were
observed at elevations $\geq 20^{\circ}$.  In all cases efforts were
made to observe fields at the highest elevations possible.

At 74~MHz scattering in the solar-wind plasma distorts radio sources viewed
through it.  In order to minimize this distortion, all observations
were made at solar elongations $\geq 60^{\circ}$.
In addition, observations were
scheduled during night-time as much as possible, because the
ionosphere is less stable during the daytime, and particularly the
morning.

\subsection{Time per Pointing}
We integrated a minimum of 75 minutes on each field in order to reach
our survey sensitivity goal of 100 mJy/beam average RMS noise level at
the field centers.  In order to improve our spatial frequency coverage, 
each field was observed at multiple hour angles by dividing the total 
75 minutes on source into three shorter observations, each about 
25 minutes, that were separated in time at least one hour.  

We used Cygnus~A (3C~405) as the sole bandpass and complex-gain calibrator 
for all observations.  When the elevation of Cygnus~A was not high enough, 
we observed Virgo~A and 3C~123 instead with the intent of using them as 
calibrators.  Experience showed, however, that the bandpasses and 
instrumental gains were stable (to within a few percent) over periods of
as long as two days.  As a result, using calibration gains derived from 
the nearest Cygnus~A observation in time (up to two days earlier or later) 
proved more reliable in our pipeline reduction than trying to use the two 
weaker calibrator sources.  Therefore, neither Virgo~A nor 3C~123 was
ever used as a calibrator.  We typically observed a calibrator for three to 
five minutes once every two hours.  We did not observe any
secondary or phase calibrators for two main reasons.  First, because
of the large primary beam size and low antenna gain at 74~MHz there are 
only a few sources in the entire sky that would be suitable.  Second,
because of the angular structure of the ionospheric fluctuations, 
gains calculated more than a few degrees away from the target source are 
not useful.

We planned 1.5 hours of time per pointing center to accommodate the 75
minutes integration on source, slewing between fields, and calibration
scans.  The slew times necessarily varied from observation to
observation; when available, extra time was used to integrate longer
and/or increase the number of scans on each pointing.  Typically,
observations were made in 6-20 hour blocks. 

\subsection{Re-Observations}

For about 65\% of the fields, the above observation strategy was
sufficient to produce maps that met our survey criteria.  However,
some fields had unacceptably high RMS noise levels after 75 minutes of
integration.  The principal causes of high map noise were RFI  or 
strong ionospheric turbulence, which can render unusable 
some fraction of the observing time on a given field.
We have re-observed most of these fields with
additional 20-25 minute scans.  The new data were then combined with
the old and a new map (usually with lower noise) was produced.  If that map
still had high noise levels, more time was scheduled.
Including time for calibration and slewing, nearly 20\% of our total
project time was used for re-observations of this nature.

\section{Data Reduction Method}
\label{reduction.sec}

\subsection{Calibration}

\subsubsection{Cygnus~A as a Calibrator}

VLSS data were calibrated using the radio galaxy Cygnus~A (3C~405).  Its 
17,000 Jy is by far the highest flux density for any non-variable object 
in the sky at 74~MHz.  (Cassiopeia~A has a similar flux density, but this is 
not constant in time, and it is a much more extended source and therefore is
fainter on the longer baselines.)  This high flux density is crucial because 
to 
function as a calibrator, a source must dominate the total flux density in its
field of view, and at 74~MHz the $11.9^{\circ}$ field typically contains 
several hundred Jy of flux density in background sources.  Another advantage
of such high flux density is that even in narrow (12~kHz) channels the source
is nearly always far stronger than any RFI.  Because of its relatively 
high declination, Cygnus~A is at an elevation of at least $30^\circ$ for 
over 10 hours per day at the VLA, a time period
which nearly always overlapped with some portion of our observing sessions.
For the few observations where it was never at high enough elevation, 
a scan from the day before or day after was used, as this was more reliable 
than using a weaker source and kept the flux scale consistent.
We observed Cygnus~A for roughly 3 minutes
every 2 hours or so while it was above an elevation of  
$30^{\circ}$.  Even at our resolution of $80''$, Cygnus~A is heavily 
resolved, and we used a pre-existing image as a calibration model
(available from {\tt http://lwa.nrl.navy.mil/tutorial}).  This 
model has been scaled to have the same total flux density as
calculated using the spectral model from \citet{baars77}, which is 17,086 Jy
at 73.8~MHz.  This defines the flux-density scale of the entire survey.
The accuracy and reliability of this flux-density scale will be discussed 
in further detail in Section~\ref{flux.scale}.

\subsubsection{Bandpass and Amplitude Gain Calibration}
\label{bpasscal}

As our observations were conducted in spectral-line mode, it was necessary to
perform a bandpass calibration.  This was done for each observation cycle 
using the existing scans of Cygnus~A and using the pre-existing Cygnus~A 
model.  A few channels at the center, known to be generally free of RFI, were 
used to normalize the bandpass and set the zero-point for the phases.  The
resulting bandpass solutions were then inspected by hand.  Occasionally one
or more antennas were not functional, and this could immediately be identified
as a bandpass that appeared pathologically shaped or that was simply random
noise.  Once identified, the data from these antennas were flagged from 
all sources for these times and the bandpasses for all antennas were 
re-calculated.

Next we performed a gain calibration, again by comparing the scans of 
Cygnus~A to its pre-existing model.  This again provided an opportunity 
to remove defective data.  We removed data showing amplitude solutions 
with a scatter among adjacent time intervals that was much greater than 
normal.  Also, typical amplitude gains at 74~MHz vary over time by about 
10\% or less, and so, if an antenna varied by much more than this, it was 
also flagged.  As we could only see these solutions during the times we 
observed Cygnus~A, we removed all data surrounding a ``bad'' scan on 
Cygnus~A.

The gain calibration produced reliable 
amplitude gains; however, the gain phases depend greatly on the location in 
the sky and therefore could not be transfered from Cygnus~A to any given 
field.  Determination of the gain phases toward the field of interest would 
be done at a later stage.  

\subsubsection{Instrumental Phase Gains}

Although we could not fully determine the phase gains simply by calibrating
to Cygnus~A, it
was necessary to estimate the instrumental contributions to the phase
as they are not corrected by the technique described later.
For any baseline, the observed phase is the sum of four components:
\begin{equation}
\label{phaseterms.eqn}
\phi\ =\ \phi_{src}\ +\ \phi_{inst}\ +\ \phi_{ion}\ +\ \phi_{noise}
\end{equation}
where $\phi_{src}$ is the phase contributed by the structure of the
source, $\phi_{inst}$ is the phase contributed by the VLA instruments, 
$\phi_{ion}$ is the phase produced by the
ionosphere and $\phi_{noise}$ is due to the noise from the Galactic
background and the thermal noise in the electronics.

Observations of Cygnus~A can be used to isolate these components and
obtain an estimate of $\phi_{inst}$.  This is because Cygnus~A is such a
strong source that it completely dominates its field of view and therefore
the overwhelming contribution to $\phi_{ion}$ comes from a single isoplanatic 
patch.  The first step in determining $\phi_{inst}$ was to observe Cygnus~A 
in a number of scans separated in time.  A standard phase-calibration 
procedure was used to estimate the antenna-based phases every 10 to 30 
seconds and a model of Cygnus~A was used to separate this from the phases 
produced by the source morphology, $\phi_{src}$.  Since $\phi_{noise}$ is 
assumed to be small and uncorrelated in time, it is assumed to
average out to a negligible level over the course of the several scans
on Cygnus~A and can be ignored in the following.
This leaves the antenna based $\phi_{inst}$ and $\phi_{ion}$ terms in
the calibration results.
The ionospheric phase is the sum of a linear gradient across the
array and higher-order terms.  The linear gradient causes a position
shift and the higher-order terms defocus the array.
It is not possible to uniquely determine a set of instrumental phases
as it is not possible to distinguish among the set of instrumental
phases plus a linear gradient across the array.
However, since the data calibration procedure described later can
determine and correct linear gradients, a set of instrumental phases
plus an arbitrary linear gradient is sufficient.

A reference time segment of well-behaved data is used to define the
instrumental plus linear phase gradient.
A calibration at a single time at which there were no higher order
ionospheric phase terms would be sufficient for calibration; however,
this is infrequent and cannot be established from measurements at a
single time.
Thus, a least squares procedure is used in which the time sequence of
calibrator results are used to fit:
\begin{enumerate}
\item Instrumental phase\\
per antenna and receiver (plus an arbitrary linear gradient).
\item Linear Gradient\\
across the array for each calibration time with respect to the reference 
time interval.
\end{enumerate}
The higher-order ionospheric phase terms are assumed uncorrelated over
the time range of the calibration data implying that their
influence will average out.
Residuals from the fitting can be used to determine when the
ionosphere is too disturbed and the higher-order terms dominate, as
well as the occasional phase jumps in the VLA electronics.
Periods of overly disturbed ionospheric conditions were flagged and the
calibration procedure repeated.
Occasionally phase jumps were discovered in the data and they were 
corrected and the calibration procedure repeated.
Once the instrumental phases are used to correct the data, it is
possible to use the data to image the sky, albeit with a time and 
position dependent systematic position offset.

\subsection{RFI Excision}
\label{flagging.sec}

\subsubsection{Need for Automated Procedure}

RFI causes excess signal to appear in the visibility data.  The best way to 
remove this is by looking at plots of phase or amplitude as a function of 
time and frequency channel and removing contaminated data by hand 
\citep{lane05}.  However, with 523 fields, each with 120 channels, 351 
baselines and both right and left circular polarizations, there was no 
practical way to perform this type of flagging by hand.  Therefore flagging 
of data contaminated with RFI was done using automated procedures.  

\subsubsection{Removing Bad Channels}
\label{comb.sec}

\begin{figure}
\begin{center}
\includegraphics[height=7in]{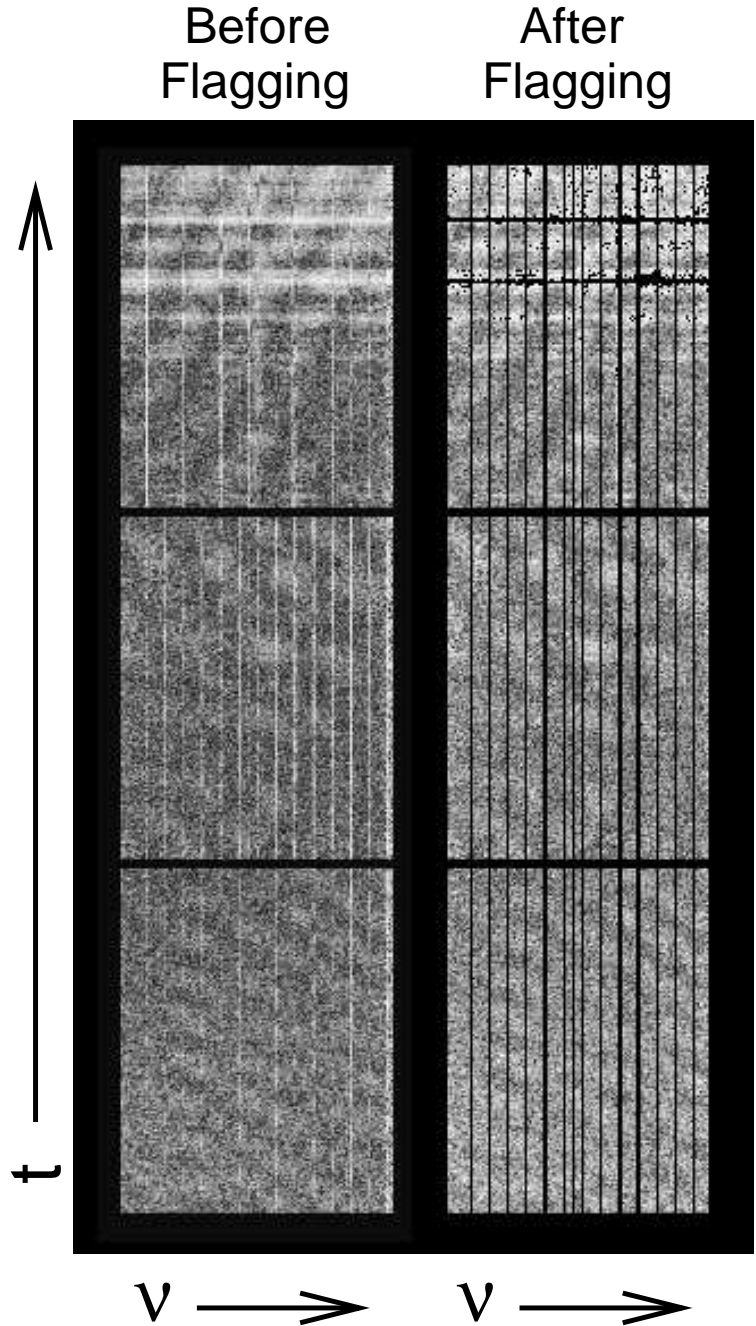}
\end{center}
\caption{\small An illustration of the automated flagging procedure.  Visibility 
amplitudes are shown for a single baseline in right circular 
polarization plotted with frequency channel along the horizontal axis and
time interval along the vertical axis.  The time is divided into three scans
of roughly 25 minutes each.  On the left are the data before flagging.  
Most of the smooth diagonal 
features are actual source structure.  The entirely vertical features 
show channels in which RFI was present for long periods of time and the 
horizontal features show time ranges in which RFI was present across a 
wide range of frequencies.  On the right are the data remaining after 
applying the automated flagging routines described in 
Section~\ref{flagging.sec}.  Most of the worst RFI is successfully removed; 
however, some amount remains in particularly bad regions.
\label{rflag.fig}}
\end{figure}

The first step in flagging was to remove channels which are known to nearly 
always be contaminated with internally produced RFI.  These channels are 
nearly equally spaced across the bandpass (seen in Figure~\ref{rflag.fig}).  
This internally generated interference is often seen in 74~MHz VLA data and 
comes from the VLA DCS system (used to send command and control data around 
the array) and results in a well-known ``100~kHz comb'' of narrow-band 
harmonics distributed across the bandpass.

\subsubsection{Clipping Ultra-High Visibilities}
\label{clip.sec}

The next step in our automated RFI-flagging procedure was to ``clip'' all 
visibilities with amplitudes above a flux level greater than what could 
conceivably come from astronomical sources in the field of view.  The 
clipping threshold was determined automatically using an algorithm that 
fit the existing $uv$-data to solve for the 
zero-spacing flux density (ie. total flux density in the field of view).  
The total flux density in a field of view had a median value of about 
330~Jy for all fields we observed, but of course certain fields with 
unusually strong sources had much higher total flux density.  
For a given field, we flagged all data points that were more than twice 
the estimated total flux density of that field.  
This typically removed about 5-10\% of the data, though for some fields 
it was as high as 20\%.

\subsubsection{Flagging by Statistical Tests}

Removing the ``comb'' and clipping removed the worst data; however, RFI 
often shows up as a 
lower-level effect in many different visibilities that form recognizable 
patterns that a human can identify, but which are below any reasonable 
clipping threshold.  This RFI usually produces excess signal in one channel 
for a long time or in many channels for a short time.  Statistical tests 
on the visibility data are necessary to identify this type of RFI.  For a 
given field, we examined the statistical properties for each baseline and 
polarization separately.  The flagging based on these statistical properties 
represents the final step in our flagging procedure.

For a given field, we examined separately each set of visibilities 
for a given baseline and polarization.  
Within this set, each visibility can be identified by its time and frequency, 
and its amplitude can be represented as a function of these, $S(t_i, \nu_j)$, 
where the time $t_i$ and frequency $\nu_j$ are those for the $i$th time 
interval and $j$th frequency respectively. 
There are 120 frequency channels kept after removing attenuated channels at 
edges of the bandpass and 450 10-second time intervals for a typical
observation of 75 minutes duration.  This results in a total of 54,000 
visibility data points.  

First we searched for individual points contaminated
by RFI.  This was done by calculating the mean and RMS values for the 
amplitudes of all 54,000 points.  We then flagged all
points having amplitudes greater than the mean amplitude plus ten times
the RMS amplitude.

The second step was to search for RFI-contaminated channels.  For this 
purpose, the time average of all data in each channel, $S_{\nu}(\nu_j)$ 
was calculated as:
\begin{equation}
S_{\nu}(\nu_j) = \frac{1}{N_{time}}\sum_{i = 1}^{N_{time}} S(t_i, \nu_j)
\end{equation}
\noindent
where $N_{time}$ is the number of 10-second time intervals.  We then 
calculated the median value of $S_{\nu}(\nu_j)$ over all $\nu_j$, which we 
call $S_{\nu,med}$.  We also define $\Delta S_{\nu}$ as the difference 
between the median and the value of $S_{\nu}(\nu_j)$ for which 25\% of the 
channels are lower (ie. the difference between the 25th and 50th 
percentiles).  For any channel whose median flux exceeded $S_{\nu,med}$ 
by more than 6$\Delta S_{\nu}$, we flagged all visibilities in that 
channel.  As RFI can only add power, we only needed to flag channels 
that are too high, rather than any that might be too low.

Finally, we searched for times in which all or most channels were 
contaminated by RFI.  This is done in nearly the same way as in step two, 
but with time and frequency exchanged.  For each time, 
all frequency channels were averaged to give:
\begin{equation}
S_{t}(t_i) = \frac{1}{N_{chan}}\sum_{j = 1}^{N_{chan}} S(t_i, \nu_j)
\end{equation}
\noindent
where $N_{chan} = 120$ is the number of channels.  We then calculated the 
median value of $S_{t}(t_i)$ over all $t_i$, which we call $S_{t,med}$. 
We define $\Delta S_{t}$, as the difference between the 25th and 50th 
percentiles of $S_{t}(t_i)$.  We flagged all times that had a median flux 
that exceeded $S_{t,med}$ by more than 4$\Delta S_{t}$.

Note from the above discussion that we flagged individual visibilities, 
channels, and time intervals based on 10$\sigma$, 6$\Delta S_{\nu}$ and 
4$\Delta S_{t}$ criteria respectively.  
These levels were determined empirically based on trial and error
in order to remove the most RFI possible without removing significant amounts
of real features.  We erred on the side of leaving in RFI rather than removing
real data, and some low-level RFI was inevitably left in much of the data.  
Again, we emphasize that the best method currently available for flagging 
RFI is to do so by eye.  
That being impossible for this survey, our goal was to removed the worst of 
the RFI, which resulted in acceptable image quality overall.

Figure~\ref{rflag.fig} shows the result of applying our flagging procedure to
a baseline with a relatively bad case of RFI contamination.  Notice that RFI
seemed to affect a set of channels in an almost equally spaced ``comb'', as 
described in Section~\ref{comb.sec}.  However, other channels were also 
contaminated with RFI and were removed.  
Also seen are horizontal features which are caused by times
during which nearly all channels were contaminated.  The smooth features, 
sometimes appearing as diagonal stripes, are from real source structure.  
As can be seen, most of the worst RFI was successfully removed; however, 
some low-level contamination remains.  

\subsection{Producing Images}
\label{imaging.sec}

\subsubsection{Channel Averaging}

After flagging, we averaged the data to reduce processing time.  The
data were averaged down to only 12 channels, the minimum we could retain 
without introducing significant bandwidth smearing into the images.  This 
resulted in channel widths of 122~kHz.  Bandwidth smearing occurs in the
radial direction with respect to the primary-beam center, and its magnitude 
is proportional to the distance from the primary-beam center.  In our case, 
a point source on the primary beam half-power circle is convolved
radially with a function 35.3$''$ wide.  This broadens the radial
width of the point-source response from 80$''$ to 87.4$''$ at
the half-power circle.  This effect is of roughly
the same magnitude as the smearing from ionospheric effects, which will 
be discussed later.

\subsubsection{Field-Based Calibration}
\label{fieldbased.sec}

Ionospheric phase errors $\phi_{ion}$ (see Equation~\ref{phaseterms.eqn})
must be removed before the (u,v)-data sets are
Fourier transformed to make images.  At frequencies significantly
higher than 74 MHz, the ionospheric phase at any instant is nearly
constant across the primary beam of a VLA antenna and can be removed
by simple antenna-based calibration.  Both ionospheric phases 
\citep{kassim93} and primary beamwidths scale as $\nu^{-1}$, so the angular 
size of the
"isoplanatic patch" over which the ionospheric phase is constant
becomes smaller than the primary beam at low frequencies.  At 74 MHz
the isoplanatic patch is significantly smaller than the VLA primary
beam, and antenna-based calibration is incapable of removing
ionospheric phase errors throughout the field of view.  Instead, we
must solve for $\phi_{ion}$ as a function of position within the field of
view.

This was done for the survey data with a method called ``Field-Based 
Calibration'' \citep{cotton04}.  Developed specifically for 74~MHz VLA data,
this method fits a time-variable phase screen over the field of view.  
Field-based calibration relies on two main assumptions.  First it assumes 
that the phase screen is the same for all antennas.  This is reasonable 
to assume for the VLA B-configuration because its maximum baseline is 11~km, 
which is small compared to the size of the isoplanatic patch at the 
altitude (about 400~km) of the maximum electron density of the
ionosphere.  This calibration method fails for arrays that are much larger 
than the isoplanatic patch projected onto the ionosphere.
The second assumption is that the spatial structure of the 
ionosphere is smooth enough that across any individual source it can be 
approximated as a simple linear gradient, which affects source images only 
by shifting their apparent sky positions.  This 
is true most of the time, but for periods of unusual ionospheric activity
this assumption no longer holds, and imaging during these times is not 
possible with this calibration method.

Field-based calibration is implemented by dividing the data into time 
intervals short enough that the ionosphere does not vary significantly, 
generally 1-2 minutes.  Within each time interval, small images are produced
of sources known to have high flux density after extrapolating from the NVSS 
with $\alpha = -0.7$.  Due to the 
short time interval these maps have high noise levels (about 1~Jy/beam), 
and typically only 5-10 sources in a given field will have high enough 
peak brightness to be clearly detected.  These sources can be used as
ionospheric phase calibrators.  We compared their apparent positions
at 74 MHz with their NVSS positions and used the offsets to determine
the phase gradients at the source positions.
A 2nd-order Zernike polynomial phase 
screen is then fitted to these phase-gradient measurements.  Higher-order 
fits are not possible because there are typically not enough detectable
calibrator sources in a field to constrain higher-order terms.  However, 
under normal ionospheric conditions, the 2nd order Zernike polynomial fit
is sufficient to remove most of the ionospheric distortions.  Residual 
ionospheric distortions will be discussed further in later sections on the
analysis of image quality.

\begin{figure}
\begin{center}
\includegraphics[width=6.5in]{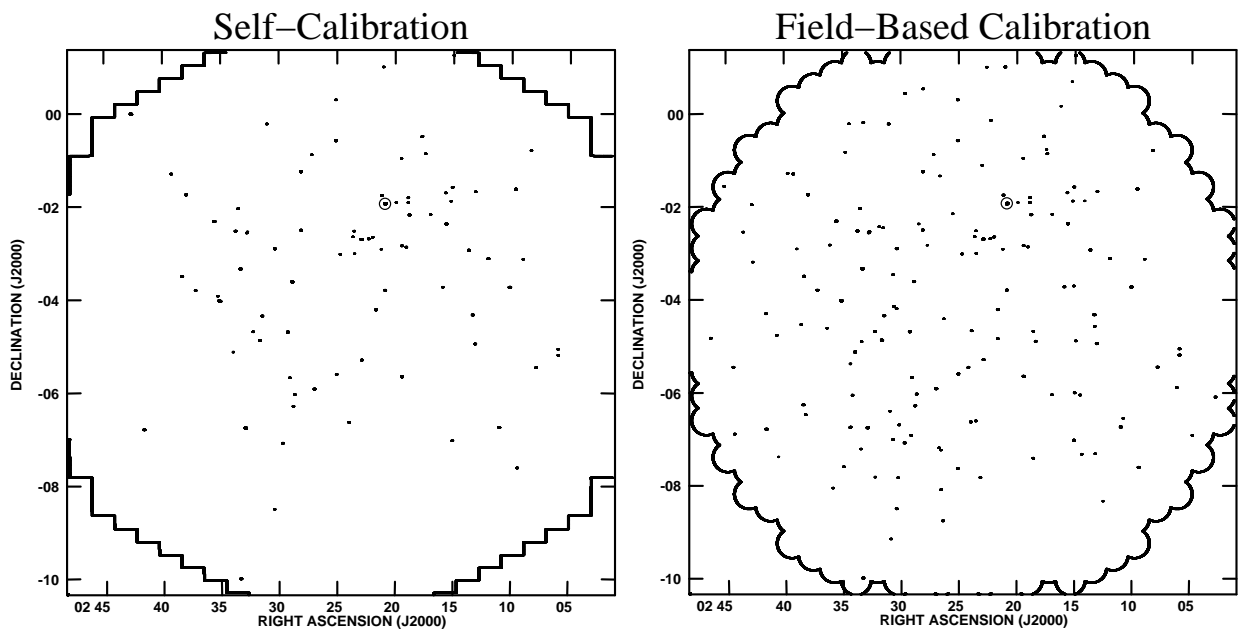}
\end{center}
\caption{
Comparison of self-calibration versus field-based calibration for the same
74~MHz data set \citep{cohen03}.  
For each method, all sources with peak brightness above
400~mJy/beam are plotted.  The circled source is 3C~63, which at 35~Jy is 
much stronger than any other source in the field of view and therefore 
dominates the self-calibration.  In the self-calibrated
image, sources far from 3C~63 suffer increased ionospheric smearing due to 
increasingly un-correlated ionospheric phases.  This causes the peak brightness
of sources to decrease, lowering the apparent source density in the image.
This problem is greatly improved in the field-based-calibrated image, which 
shows roughly uniform source density throughout the field of view.  
\label{4A.maps.fig}}
\end{figure}

The improvement in image quality of field-based calibration compared to 
self-calibration can be seen in the source maps of Figure~\ref{4A.maps.fig},
which was taken from \citet{cohen03}.  For each map, the dots represent all 
sources in the image with apparent peak brightnesses of 400 mJy/beam or 
higher.  This 74~MHz VLA data set was observed in the A-configuration, for 
which ionospheric effects are much more pronounced than in the smaller 
B- and BnA-configurations used for the VLSS, and therefore shows a very 
marked contrast
between the resulting image quality of the two methods.  The flux density 
in this field of view is dominated by 3C~63, which is circled, causing 
self-calibration to solve for ionospheric phase at that location and 
subtract this phase over the entire field of view.  At increasing angular
distances from 3C~63, the ionospheric phases have less correlation with 
these solutions, and sources appear to have position shifts that vary 
in time.  As the image is produced by averaging the data over time, this 
causes source smearing which reduces the apparent peak brightness and thus 
the apparent source density.  This is clearly seen in the source map for
the self-calibrated image, which shows a declining source density with 
increasing angular distance from 3C~63.  In contrast, the field-based 
image has a roughly uniform source density throughout which only decreases
at the edges because of primary-beam attenuation.

\subsubsection{Wide-Field Imaging}

After determining the ionospheric phases, the next step is Fourier inversion 
into the image plane.  Normally, this is done
with a two-dimensional Fourier inversion of the visibilities by projecting
the baseline vectors onto the $uv$-plane perpendicular to the line of sight.
However, this is an approximation that is only valid for small fields of 
view in which $\theta^2\,w_{max} \ll 1$, where $w_{max}$ is the maximum  
baseline component parallel to the line of sight for all visibilities in 
the data set measured in wavelengths, and $\theta$ is the angular size of 
the imaged region in radians.  
The 74~MHz field of view is far too large for 
this approximation to be valid.  Therefore, conversion to the image plane
was done with polyhedral imaging
that divides the field of view into smaller plane images (facets)
inside of which the two-dimensional approximation 
is valid \citep{cornwell92, perley99}.
The facet size was set automatically for each data set such that 
$\theta^2\,w_{max} = 0.01$, resulting in facets that were typically about 
$1^\circ$ in size or less.  Depending on the facet size, each field was 
covered by between roughly 250 and 1500 facets.  Additional facets were 
placed at the known locations of very strong sources outside of the field 
of view so that their deconvolution could reduce the effects of their 
sidelobes within the field of view.

The ionospheric phase screens, determined in Section~\ref{fieldbased.sec}, 
were then applied by using the appropriate $\phi_{ion.}$ for each facet
as determined by the location of that facet within the phase screen.  
The facets
were deconvolved using the CLEAN algorithm with a constant circular 
restoring beam with $80''$ FWHM.  The facets were then combined into a single
calibrated and astrometrically corrected image of the entire primary beam 
region.

\subsubsection{Removal of Strong Outlier Sources}

In some cases the location of a field near an unusually strong source could 
cause high sidelobes in the field of view, which greatly increased the
overall noise level.  In bad cases it could even cause the ionospheric 
calibration to fail altogether.  In these cases we first self-calibrated 
the data to the problematic source, then mapped that source and subtracted
it from the $uv$-data, and then reversed the calibration before finally 
proceeding with the imaging as normal.  This procedure is generally known
as ``peeling''.  This greatly improved the image
quality for many fields, although for extremely strong sources, the 
surrounding fields still have higher noise levels than average 
because the source subtraction is not perfect and leaves residual errors.

\subsubsection{Corrections of Residual Ionospheric Calibration Errors}

\begin{figure}
\begin{center}
\includegraphics[height=7.25in]{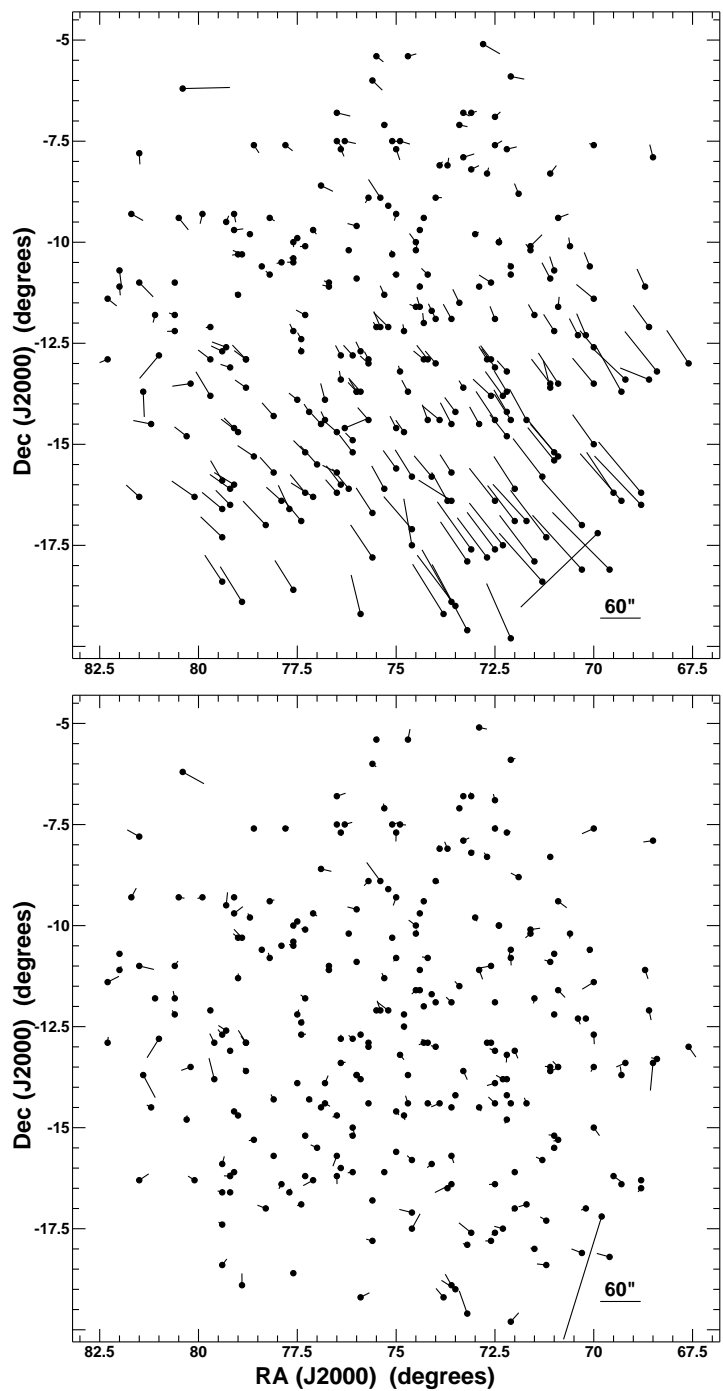}
\end{center}
\caption{
Source shifts relative to the NVSS positions for a VLSS field that had one 
of the worst cases of ionospheric calibration errors.  The 
dots represent the VLSS source positions and the lines represent the distance 
and angle to the NVSS position.  The shift magnitudes are magnified according
to the scale shown at the lower right in order to be visible.  The 
upper plot is before the corrections and the lower plot is after.  Individual 
sources may be truly shifted from the NVSS positions due to actual differences
in source centroids between  1400 and 74 MHz, however any true shifts should
be random and not correlated with nearby sources.  
\label{shift.fig}}
\end{figure}

After examining many field images, we discovered that some fraction of them 
contained residual ionospheric errors that shifted sources
from their NVSS positions.  There are 
three potential reasons for VLSS source positions to be different from their
NVSS positions: (1) source fitting errors caused by map noise, (2) source 
centroids being truly 
different at 74 and 1400~MHz because of spectral variations across the source
and (3) the VLSS image at the source 
location is shifted due to ionospheric calibration errors.  
Reasons (1) and (2) will cause (usually small) random shifts which are not 
correlated with the shifts of nearby sources.  These shifts are expected and
do not require correction.  However reason (3) can cause all the sources in a 
given region to be shifted by nearly the same magnitude and direction, thereby
making the image astrometrically incorrect in that region.  The 
fact that this was seen in some 
images (Figure \ref{shift.fig}, top plot) indicates that ionospheric
errors were significantly affecting some of the data.

Ionospheric calibration errors can be separated into errors that are
time-dependent and those that are time-independent during the observation.  
Time-dependent errors could be caused by ionospheric variations that are too 
complex in 
time or space to be modeled by our field-based calibration scheme.  There is 
nothing that can be done about these errors because they must be corrected 
individually for each solution time interval.  Each such time interval,
by itself, does not produce an image deep enough to detect enough sources
to apply a more sophisticated ionospheric model than was originally used.  

The main causes of time-independent errors are that one or more of the 
ionospheric calibrators used has a significantly different centroid at 
74~MHz than it does at 1.4~GHz and that the available calibrators are 
distributed in the field of view in a way that doesn't allow for a realistic
solution everywhere in the field.  These problems do not vary with time and 
cause the same calibration error for each time interval.
Unlike time-dependent errors, time-independent errors can be further 
investigated because one can use the data from all time intervals combined, 
in which generally at least 100 sources are detected in the field of view.  

Therefore we implemented an image correction algorithm designed to 
remove these time-independent ionospheric calibration errors.  
We compared the positions of sources in the
final 74~MHz image with NVSS positions and fit a up to a 4th-order Zernike
polynomial correction, rather than the 2nd-order correction used for
the initial phase calibration.  The higher-order correction is possible 
because of the larger number of available calibrators in the integrated
image, rather than the 2-minute snapshots.  The large number of 
available calibrators also allows the algorithm to remove 
individual sources that give a bad fit to the correction models, 
making the corrections much more robust against sources with true differences
between centroid positions at the two frequencies.  The newly modeled 
``phase screen'' was applied in the image plane by stretching and then 
re-gridding the image.
Although not all fields had significant calibration errors, to be thorough 
and consistent, we applied this correction algorithm to all fields.

\begin{figure}
\begin{center}
\plotone{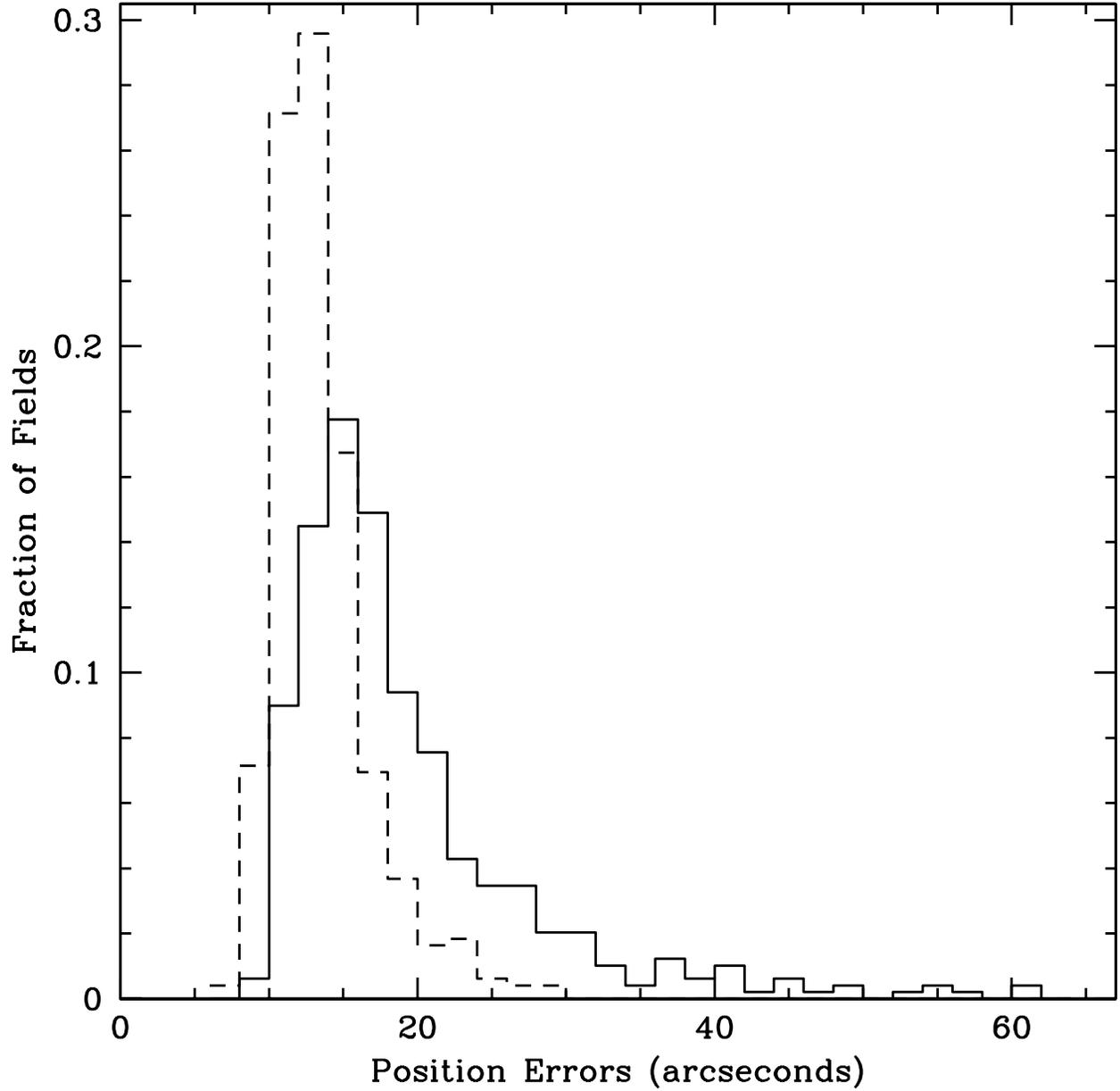}
\end{center}
\caption{
Histogram of position errors (for the 85th percentile of all sources) for 
all fields before correction (solid line) and after correction 
(dashed line).  Before correction 10\% of fields had position errors 
greater than $30''$, with some over $60''$.  After 
correction, no field had position errors more than 
$30''$, and about 95\% have position errors parameter less than
$20''$, or 1/4 of a beam width.
\label{shift.hist.fig}}
\end{figure}

The resulting maps show a nearly complete correction of correlated position 
offsets between VLSS and NVSS as can be seen by comparing the before and after
images of Figure \ref{shift.fig}.  
We defined the position error of a field as the 85th percentile in
distribution of VLSS-NVSS source offsets in the field.  The 85th
percentile was chosen to catch ionospheric errors affecting a small
portion of a field without being sensitive to a single source having a
large centroid shift between 74 and 1400 MHz.  For each field, we performed
three fits using a 2nd, 3rd and 4th-order polynomial correction.  The fit that
produced the lowest position errors (usually, but not always the 4th-order 
fit) was used for the ``final'' image.  Figure \ref{shift.hist.fig} shows the
overall improvement in the position errors of all fields.  Before
correction, 10\% of the fields had position errors greater than $30''$, with
some over 60$''$.  After correction, no field had a position error
greater than $30''$ and about 95\% had position errors less than $20''$, 
or 1/4 of the beamwidth.  Thus the entire survey is accurate to 
within $30''$.  Figure \ref{shift.hist.fig} 
also shows that almost no fields have position errors less 
than $10''$.  Position uncertainties caused by map noise probably account 
for this 10$''$ minimum.

\subsubsection{Image combining}

Individual field maps at this point were still not corrected for the 
attenuation of the primary-beam pattern, and therefore their sensitivity 
varied; it was highest at the field centers and tapered off with distance 
proportional to the primary-beam attenuation.  The individual field maps
were each truncated at a radius of $6.2^{\circ}$, which is slightly larger
than the half-power radius.  In order to produce images 
of nearly uniform sensitivity, the individual fields were combined using 
the method of \citet{condon98} which corrects for the primary-beam 
attenuation while co-adding data at each point in the final image 
weighted proportionally to the inverse square of the estimated RMS noise 
level in each contributing image.  The RMS noise levels were estimated 
by measuring the noise levels at the centers of the fields and assuming that
they increased with radius from the center with inverse proportion to the 
primary-beam attenuation.

An overlapping grid of square images with 2048 $25''$ pixels on a side 
{$\approx 14^{\circ}$} was produced to cover the entire survey region.   
This grid of combined images comprises the principal data product of the 
survey.   This grid of
images was used to produce all sub-images and the source catalog.

\section{Survey Assessment}
\subsection{Sky Area Imaged}

\begin{figure}
\plotone{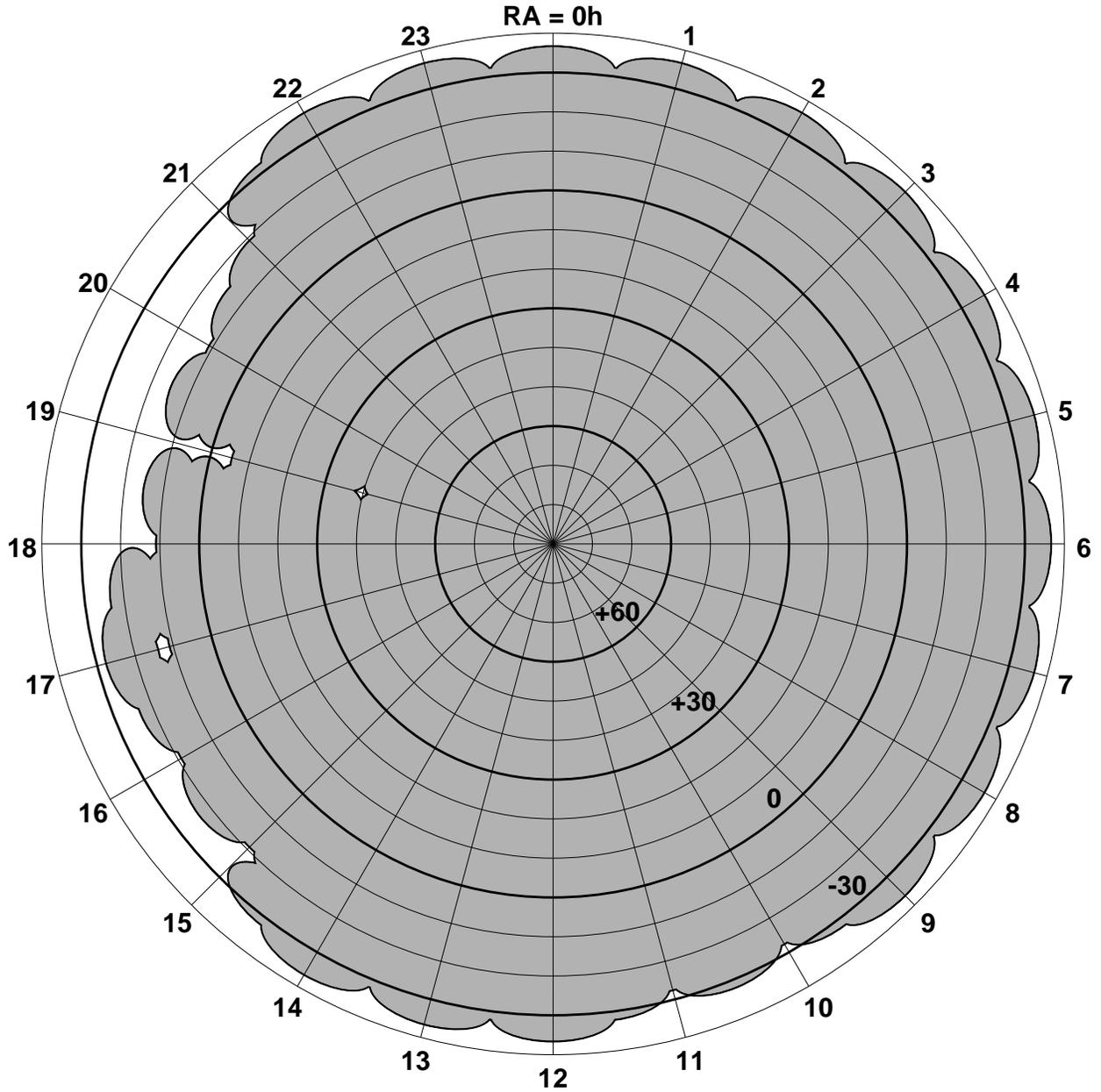}
\caption{Region of sky currently imaged by the VLSS project.  Currently 
about 95\% of the sky above $\delta > -30^{\circ}$ is now covered, and 
further observations are planned to fully complete this region.
\label{coverage.map.fig}}
\end{figure}

We have now observed, reduced, and publicly released the images and 
catalogs for the region of the sky shown in Figure \ref{coverage.map.fig}.  
This region covers a total of 9.35 sr, about 95\% of the intended survey 
region $\delta > -30^{\circ}$.
Observing time for the observations needed to image 
the remaining area have been allocated.  The full survey data will be 
released after this is completed and combined with the existing data.

\subsection{Sensitivity Achieved}
\label{sensitivity.sec}

\begin{figure}
\plotone{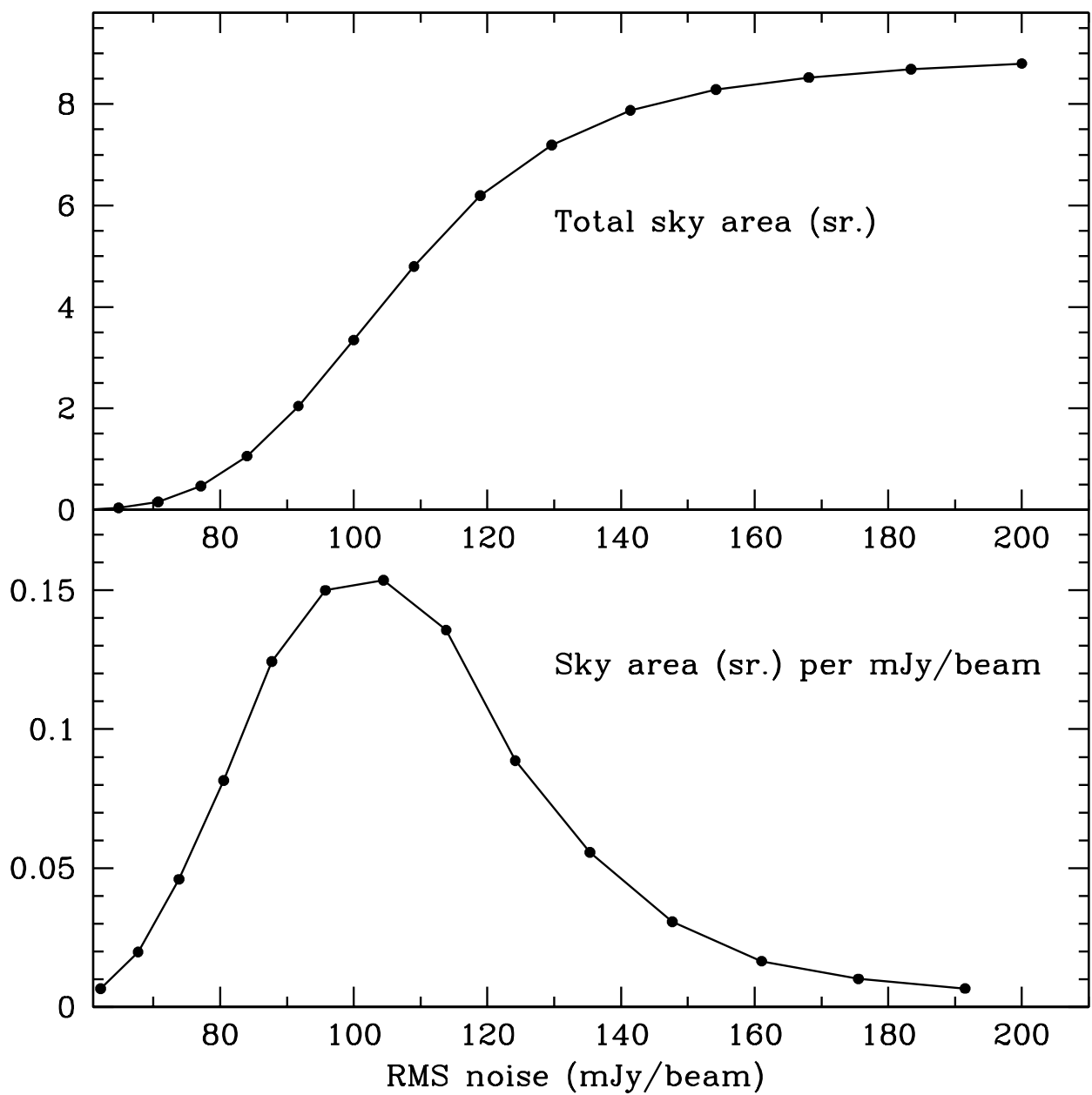}
\caption{
Top: Total sky area (y-axis) at or below a given RMS noise level (x-axis). 
Bottom: Differential sky area (y-axis) at a given RMS noise level (x-axis).
The median RMS noise level is 108~mJy/beam.
The extended ``tail'' at higher noise levels is due 
mostly to regions around very bright sources (such as Cygnus~A, Casseopeia~A, 
Virgo~A, etc.), regions of sky with high sky temperature like the Galactic 
plane, and regions at the edges of the mosaic maps which have no 
overlap with other fields.
\label{area.stats.fig}}
\end{figure}

For the sky area shown in Figure \ref{coverage.map.fig} we achieved
a median RMS noise level of 108 mJy/beam.  However, some
regions of the sky had significantly higher noise levels.  This can be 
seen in Figure \ref{area.stats.fig} which quantifies how much sky was 
observed at each RMS noise level.  In the lower plot, one can see that 
while most of the sky area has now been observed at $100\pm30$mJy/beam, 
there is a significant ``tail'' extending to much higher noise levels.
The main causes for this are: (1) regions near very strong            
sources in which
the RMS noise level is dynamic-range limited, (2) regions at the edges
of fields with no neighboring field in which the primary beam shape raises
the RMS noise level (3) field located in high-sky-temperature regions in the 
Galactic plane near the Galactic center, and (4) fields that were affected 
by unusually bad ionospheric conditions or RFI conditions.  We can do
nothing about causes (1) and (2), but we intend to re-observe fields affected 
by causes (3) and (4) to reduce this effect in the next data release.

The sensitivity of an image produced with the 74~MHz VLA is generally not
thermal noise limited.  For a typical VLSS observation of 75 minutes with 
26 working antennas, the theoretical RMS noise is 35~mJy/beam for a system 
temperature $T_{sys}$ of 1500K and aperture efficiency of 0.15.  
This value of $T_{sys}$,
which is dominated by the sky temperature $T_{sky}$, is typical for the 
sky far from the Galactic plane, but can be up to twice that on the 
Galactic plane and even higher still near the Galactic center.  However, 
the map noise is dominated not by $T_{sys}$, but by sidelobe confusion 
from the thousands of sources inside and outside the primary beam area, 
an effect that is accentuated by the poor forward gain and commensurate high
sidelobe levels of the 74 MHz primary beam.  
This is why most fields have noise levels that are between two to four times 
the thermal noise.  

\begin{figure}
\begin{center}
\includegraphics[height=7.5in]{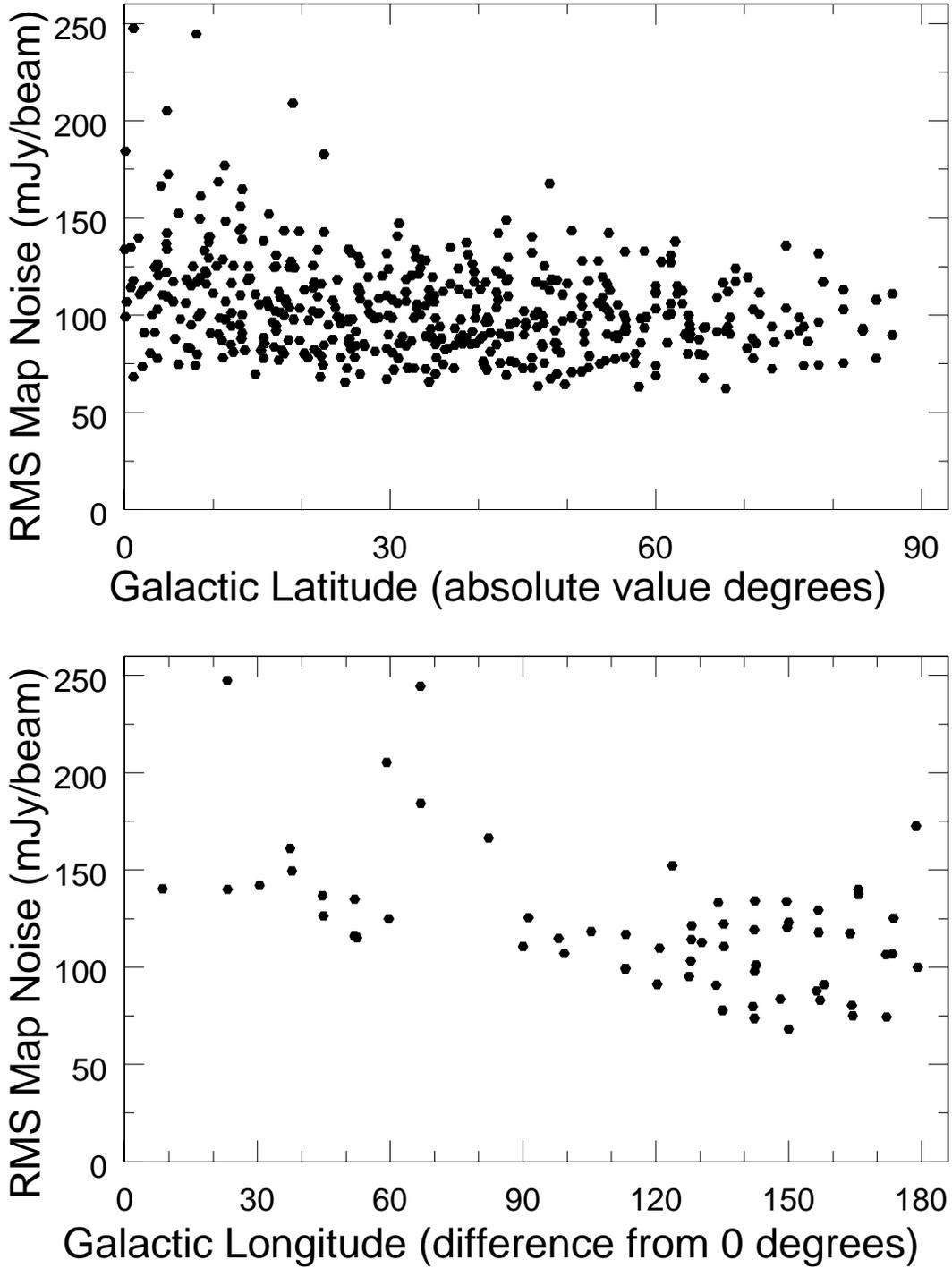}
\end{center}
\caption{
Top: RMS map noise for each field plotted against the absolute value of 
the Galactic latitude of the field center.
Bottom: RMS map noise for each field located within $10^\circ$ of the 
Galactic plane plotted against the Galactic longitude of the field center
in terms of degrees from the Galactic center.  For both plots, fields located
close to extremely strong sources such as Cygnus~A were removed.
\label{skynoise.fig}}
\end{figure}

While $T_{sky}$ is not the dominant factor producing map noise, the fact
that it varies greatly over the survey region makes it instructive to look 
for correlations between map noise and sky position.  
The top graph of Figure~\ref{skynoise.fig} shows the map noise for 
each field as a function of Galactic latitude for all fields except those 
very close to extremely strong sources such as Cygnus~A.  This plot 
indicates little
if any dependence of map noise on Galactic latitude except for a slight 
increase very close to zero latitude.  Images in the  Galactic plane region
are investigated further in the bottom graph of Figure~\ref{skynoise.fig}
which plots the map noise of all fields within $10^\circ$ of the Galactic
plane as a function of Galactic longitude.  This plot shows a clear increase
of at least 50\% in the average map noise toward the Galactic center.  This 
shows that while $T_{sky}$ is generally not a determining factor in the 
noise levels for most of the sky, it is high enough to increase map noise in 
the small region of the sky on the Galactic plane and near the Galactic 
center.

\subsection{Image Quality}

\begin{figure}
\begin{center}
\plotone{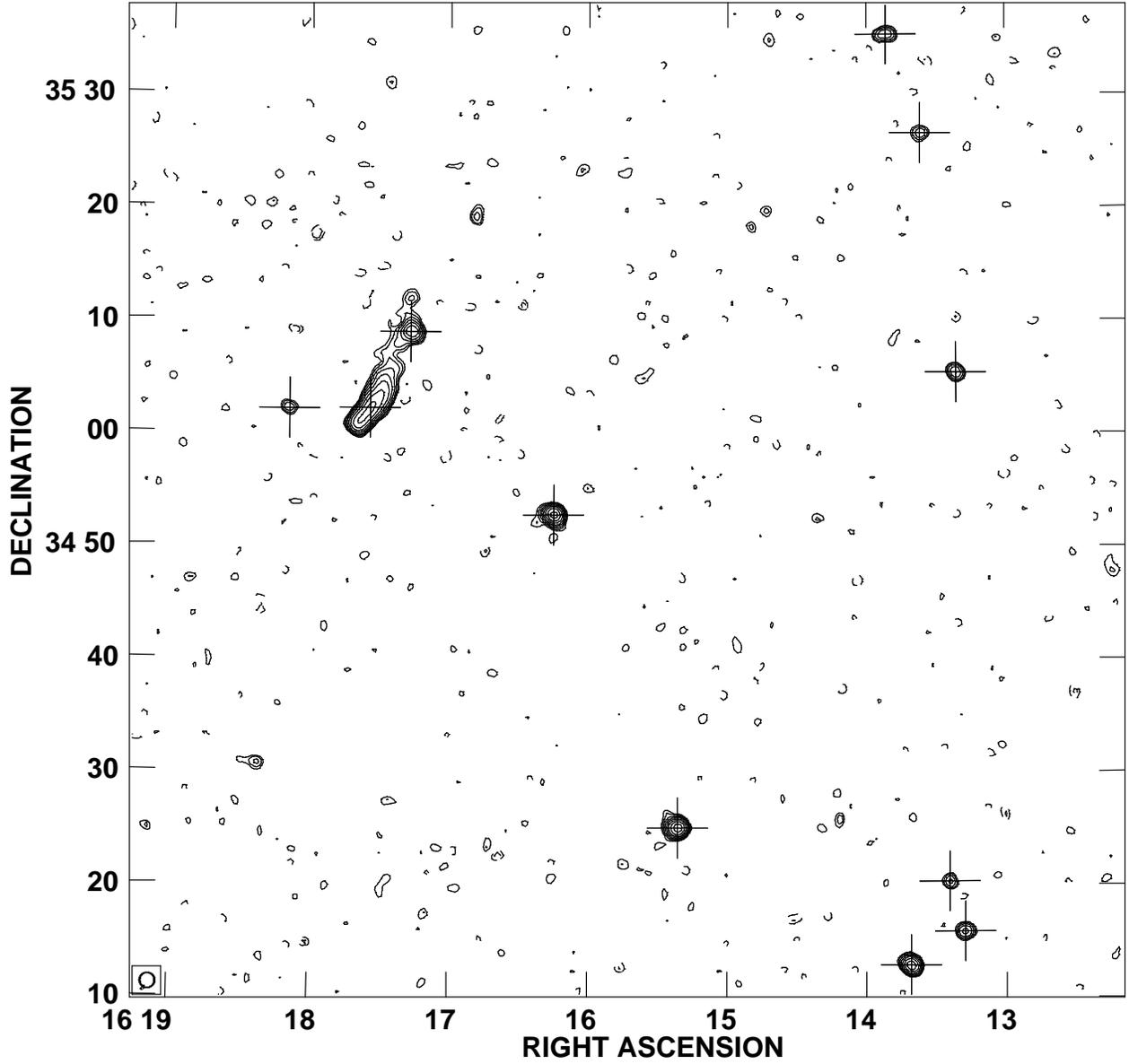}
\end{center}
\caption{
Contour plot of a sample sub-image from the VLSS.  The peak 
flux density is 2.57 Jy/beam and the overall RMS noise level is 
74.9~mJy/beam.  Contours begin at 2.5 times the RMS noise level 
(187.25~mJy/beam) and 
increase by factors of $\sqrt{2}$.  The crosses indicate the locations 
of $5\sigma$ source detections which are included in the source catalog.
\label{sample.fig}}
\end{figure}

A sample VLSS sub-image is shown in Figure \ref{sample.fig}.  The crosses 
indicate identified sources that were included in the VLSS catalog which will
be described in Section \ref{catalog.sec}.

\begin{figure}
\begin{center}
\includegraphics[height=8in]{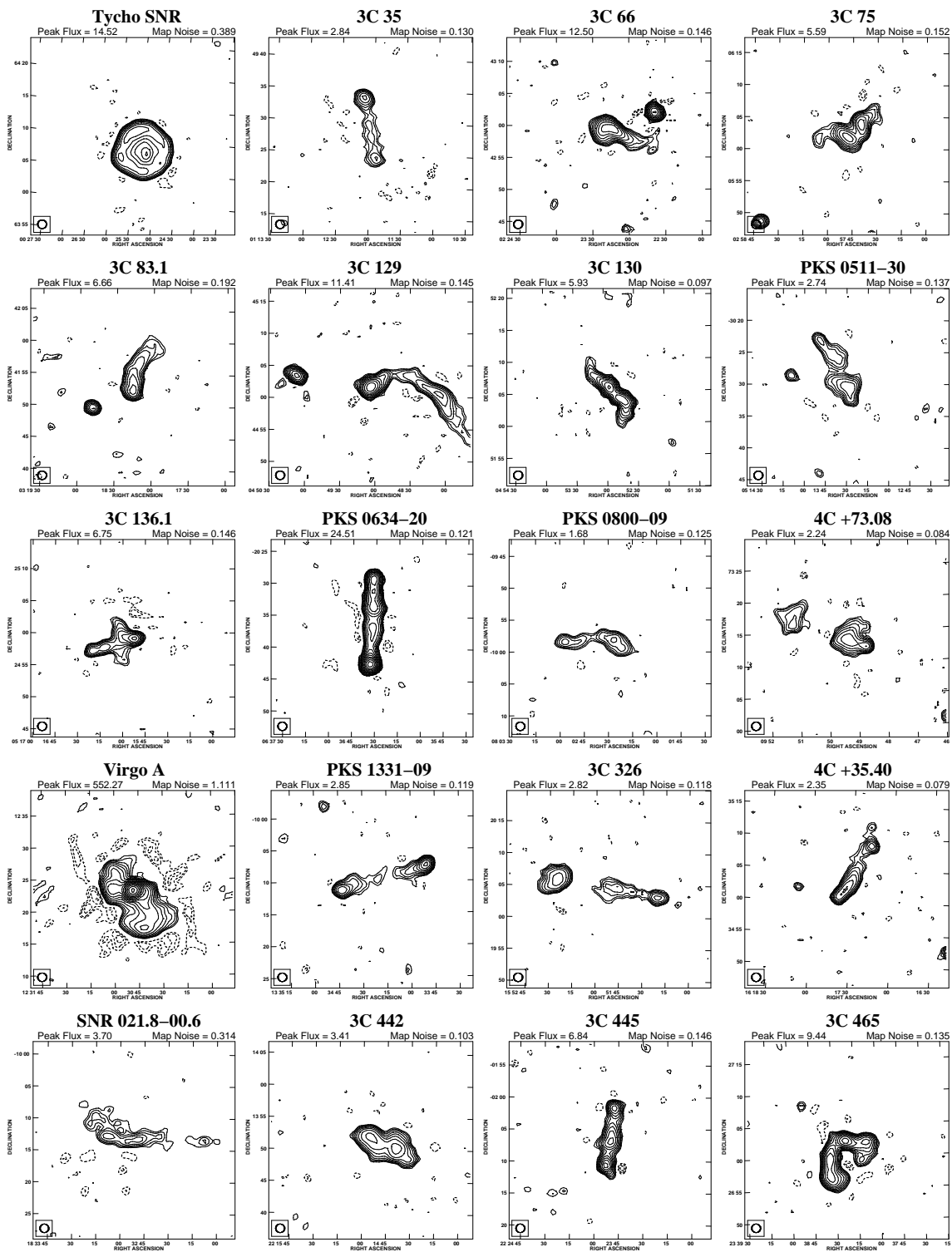}
\end{center}
\caption{
Contour plots of some of the larger sources observed by the VLSS.  Peak 
flux density and map noise are given in Jy/beam.  Contours begin at 
2.5 times the map noise and increase by factors of $\sqrt{2}$.  Sources
are each labeled by their common radio names.
\label{gallery.fig}}
\end{figure}

With the VLSS resolution of $80''$ most radio sources
are either unresolved or just slightly resolved.  However, given the large 
sky area and number of sources identified, a sizable number of very large
and resolved sources have been found.  Figure \ref{gallery.fig} shows a 
sample of some of the largest sources found in the survey.  Many of these are
well-known objects and we have labeled them by their common radio names.

\section{Extracting the Catalog}
\label{catalog.sec}

\subsection{Source Finding}

In order to produce a catalog of sources we used the same Gaussian-fitting
program used by the NVSS catalog \citep{condon98}.  This algorithm identifies 
each ``island'' of high brightness in an image that is above a specified 
threshold.  The threshold we
used was that an island had to have a peak brightness of at least 4.5 times
the local RMS noise in a square 100 pixels on a side centered on that island.  
Each island was fit to a model of up to four Gaussian peaks, which was 
generally sufficient to accurately model most source structures.  Each 
catalog entry is a single Gaussian peak.  If an ``island'' was fitted by 
multiple Gaussian peaks, they were cataloged as separate sources.  Therefore, 
a catalog ``source'' is simply a bright region that was fit by a Gaussian, 
but may actually be just one component of an astronomical source; for 
example, one lobe of a double source.  After 
fitting, only sources with peak brightnesses of 5 times the local RMS noise 
level ($\sigma$) or greater were kept in the catalog.

The result was a list of Gaussian fits, each described by six parameters:

\begin{itemize}
\item $\alpha$ = right ascension (J2000)
\item $\delta$ = declination (J2000)
\item $I_p$ = peak brightness
\item $\theta_M$ = major axis FWHM
\item $\theta_m$ = minor axis FWHM
\item $\phi_{PA}$ = Position Angle (east of north)
\end{itemize}

\noindent
These are all free parameters fit to each source, the only constraint being 
that $\theta_M$ and $\theta_m$ are required to be larger than 
$\theta_b = 80''$, the size of the restoring beam.  

\subsection{False Detections}
\label{falsesources.sec}

The $5\sigma$ cutoff for source detection should eliminate virtually all 
false source detections for the case of Gaussian map noise.  However the noise
in the VLSS images is not always Gaussian, mainly because of sidelobes from 
incompletely deconvolved sources.  Therefore in the vicinity of sources with
very high peak brightness, we apply a stricter criterion for source 
detection of $6\sigma$ peak brightness.  This ``vicinity'' was a circular 
region around each 
source with $I_p > 12\textrm{Jy/beam}$ with radius, $\theta_r$, that 
varied according to the measured peak brightness, $I_p$, as follows:

\begin{equation}
\theta_r = (1^{\circ})\sqrt{\frac{I_p}{60\,\textrm{Jy/beam}}}
\end{equation}

\noindent
up to a maximum of $6^{\circ}$.  The parameters of this equation were 
adjusted empirically in order to remove the most false sources without 
removing significant numbers of real sources.  
This removed most but not all source detections that upon visual inspection 
of the maps were clearly sidelobes of very strong sources.  In all, 549 
sources were removed in this manner.  Of these, 263 (or 48\%) had no NVSS
counterpart within $60''$, whereas for all sources, fewer than 1\% have no 
NVSS counterpart within $60''$.  While we cannot use NVSS counterparts to 
determine if individual sources are real, they are useful in comparing the 
reliability of large sets of sources.  Based on this comparison it is clear
that these 549 ``sources'' were not reliable enough to keep in the survey.  
However, 286 of these sources did have counterparts, and assuming most of 
these are real it is clear that real sources were removed from the catalog.  
Thus we increased the reliability while decreasing the completeness of the 
catalog in the vicinity of sources with very high peak brightnesses.

\subsection{Derived Parameters}

The six parameters from each Gaussian fit were used along with knowledge of
the restoring beam size, always $\theta_b = 80''$, and the local RMS map 
noise, $\sigma$, to produce the derived parameters for each source that we 
present in the VLSS catalog.  Three of the derived parameters are the same as
the fitted parameters: $\alpha$, $\delta$ and $\phi_{PA}$.  However, instead 
of reporting the peak brightness, $I_p$, we instead give the 
integrated flux density $S_i$.  Also, instead of reporting the fitted source 
sizes, $\theta_M$ and $\theta_m$, we use our knowledge of the restoring beam 
size to calculate deconvolved source sizes, $\phi_M$ and $\phi_m$.  In Section 
\ref{derivations}, we describe in detail how these derived parameters and 
their errors are determined.

\section{Derivation of Source Parameters and Their Accuracy}
\label{derivations}

\subsection{Source Positions}
\label{dpos}

The source coordinates, $\alpha$ and $\delta$, are those of the fitted 
Gaussian.  The largest contributors to their RMS errors, $\sigma_\alpha$ 
and $\sigma_\delta$, are errors in the ionospheric phase calibration and 
fitting errors caused by map noise.  Thus the total position errors are 
quadratic sums of two sources of error as are given by the following formulas:
\begin{mathletters}
\begin{eqnarray}
\label{dpos.alpha.eqn}
\sigma_\alpha^2 = \sigma_{\alpha,fit}^2 + \sigma_{\alpha,cal}^2 \\
\label{dpos.delta.eqn}
\sigma_\delta^2 = \sigma_{\delta,fit}^2 + \sigma_{\delta,cal}^2
\end{eqnarray}
\end{mathletters}
where $\sigma_{\alpha,cal}$ and $\sigma_{\delta,cal}$ are the position errors
due to calibration errors and $\sigma_{\alpha,fit}$ and $\sigma_{\delta,fit}$ 
are the fitting errors.

\subsubsection{Position Errors Caused by Calibration Errors}
\label{cal.pos.errors}
Calibration errors affect the positions of all sources, while 
map noise causes position errors that are inversely proportional to source 
flux density.
Therefore we can isolate the calibration errors by focusing on 
sources that are strong enough that the fitting errors are negligible in 
comparison.  However, most of the strongest sources were also used as 
calibrators during field-based calibration.  A source used as a 
calibrator could have a smaller calibration-induced position error than 
a typical source.  Therefore, we restrict our sample to sources with peak
brightness below 3.5 Jy/beam, which is the threshold below which calibrator
sources are rejected.  This will ensure that the sample remains unbiased.

We can estimate the position error of a source by measuring the difference
between its VLSS and NVSS positions.  For most sources, the NVSS position 
errors will be much smaller than 
the VLSS position errors because: (1) the NVSS resolution is nearly twice that 
of the VLSS ($45''$ versus $80''$), (2) virtually all sources are 
detected in NVSS at a much higher signal-to-noise ratio than in the VLSS
and (3) NVSS calibration errors are known to be only about 0.5$''$ because 
of smaller ionospheric effects at the higher observing frequency.
Thus the difference between the NVSS and VLSS positions can be assumed to be 
dominated by the 
VLSS position error.  However, there are exceptions to this for sources in 
which the source centroid is in fact different at 74~MHz than at 1.4~GHz  
because of a spatially varying spectral index throughout the source (i.e. 
the core versus a lobe of a radio galaxy).  This effect may cause a small 
contamination when calculating average position offsets for large numbers of 
sources.  However, it can only increase the average offset, and so if 
anything this method of estimating the position errors will be an 
overestimate, and therefore a more conservative error estimate.

To examine position errors of strong sources, we used only 
VLSS sources that have (1) peak brightness less than
3.5 Jy/beam (to avoid calibrator sources), (2) a detection level of at least
$30\sigma$, (3) a bright ($I_p > 50$ mJy/beam) NVSS 
counterpart within 120$''$, (4) no other counterpart within 120$''$,
and (5) fitted major axes less than 120$''$.   
The fourth and fifth criteria were included to remove fitting errors caused 
by very large or multiple-component sources having highly non-Gaussian 
shapes and the increased likelihood of having real source centroid shifts 
between the two frequencies.  There were 866 sources that met these criteria.

\begin{figure}
\plotone{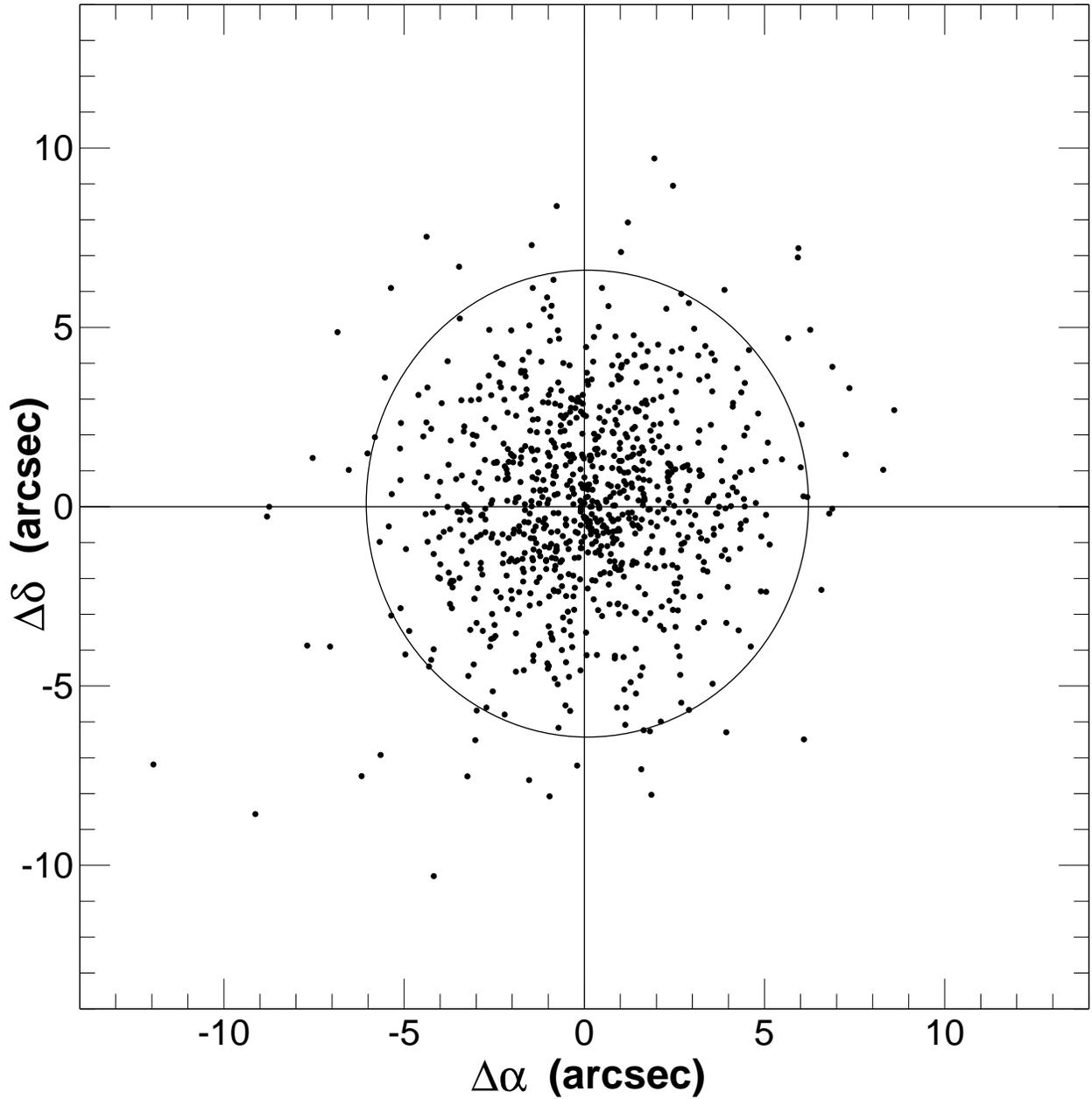}
\caption{
Source offsets of VLSS sources from their NVSS locations for the 866 sources 
strong enough that errors due to map noise are negligible, yet below the peak
brightness threshold above which they might be used as calibrators.  
The ellipse represents the 90\% confidence region based on the fitted mean 
and RMS of these offsets.
\label{dpos.strong.fig}}
\end{figure}

We then calculated the VLSS-NVSS position differences for each of these 
sources, which are plotted in Figure \ref{dpos.strong.fig}.  The results 
yielded average offset of 
$\Delta\alpha_{mean} = 0.08''$ and $\Delta\delta_{mean} = 0.08''$ and 
RMS deviation about mean values of 
$\Delta\alpha_{RMS} = 2.86''$ and $\Delta\delta_{RMS} = 3.03''$.  Because
we couldn't use the very strongest sources because they might have been 
used as calibrators, there is some level of contamination in these figures
caused by map noise errors.  For example, a point source detected at the 
30$\sigma$ level (the lowest allowed in the sample) should have a position
error in both dimensions of 1.35$''$, which subtracted in quadrature would
reduce $\Delta\alpha_{RMS}$ from  $2.86''$ to $2.52''$ and 
$\Delta\delta_{RMS}$ from $3.03''$ to $2.71''$.  However, this is only about 
a 10\% reduction and even this is only for sources at the minimum 
signal-to-noise ratio in the sample.  Therefore, while these average offsets 
could be slightly overestimated, we will proceed with them as 
conservative error estimates.  Based 
on these values, we adjusted the VLSS source positions by subtracting 
the mean offset values and set $\sigma_{\alpha,cal} = 2.86''$ and 
$\sigma_{\delta,cal} = 3.03''$ in Equations \ref{dpos.alpha.eqn} and 
\ref{dpos.delta.eqn}.  

\subsubsection{Position Errors Caused by Map Noise}

Errors caused by Gaussian fitting in the presence of map noise depend on 
the signal-to-noise ratio of the detection, $I_p /\sigma$.  However, 
\citet{condon97} found that in practice errors were more reliably predicted 
by using the following ``effective'' signal-to-noise ratio:
\begin{equation}
\label{SNR}
\rho^2 = \frac{\theta_M\theta_m}{4\theta_N^2}
\left[ 1 + \left( \frac{\theta_N}{\theta_M} \right)^2 \right]^{\alpha_M}
\left[ 1 + \left( \frac{\theta_N}{\theta_m} \right)^2 \right]^{\alpha_m}
\frac{I_p^2}{\sigma^2}
\end{equation}
for which $\theta_N$ is the FWHM of the Gaussian noise correlation function
which we can take to be the restoring beam size $\theta_b$.
The exponents $\alpha_M$ and $\alpha_m$, used for the terms adjusting for 
the major and minor axes respectively, differ depending on 
the error parameter being calculated.  

The position uncertainties are therefore given by:
\begin{mathletters}
\begin{eqnarray}
\label{cal.alpha.eqn}
\sigma_M^2 = \frac{\theta_M^2}{(4 \ln 2)\rho^2} \\
\sigma_m^2 = \frac{\theta_m^2}{(4 \ln 2)\rho^2}
\end{eqnarray}
\end{mathletters}
\noindent
where $\sigma_M$ and $\sigma_m$ are the position errors along the 
major and minor axes respectively.  The exponents used for calculating $\rho$  
from Equation \ref{SNR} are $\alpha_M = 5/2$ and $\alpha_m = 1/2$ for 
calculating $\sigma_M$, and $\alpha_M = 1/2$ and $\alpha_m = 5/2$ for
calculating $\sigma_m$ \citep{condon97}.

To get the source-fitting position errors along the 
right ascension and declination axes as needed for Equations 
\ref{dpos.alpha.eqn} and \ref{dpos.delta.eqn}, we have:
\begin{mathletters}
\begin{eqnarray}
\sigma_{\alpha,fit}^2 = \sigma_M^2\sin^2(\phi_{PA}) 
+ \sigma_m^2\cos^2(\phi_{PA}) \\
\sigma_{\delta,fit}^2 = \sigma_M^2\cos^2(\phi_{PA}) 
+ \sigma_m^2\sin^2(\phi_{PA}) 
\end{eqnarray}
\end{mathletters}
\noindent
These errors are then added
in quadrature to the calibration errors that were determined in Section 
\ref{cal.pos.errors} to produce the total position errors that we include
in the VLSS source catalog.  

\begin{figure}
\plotone{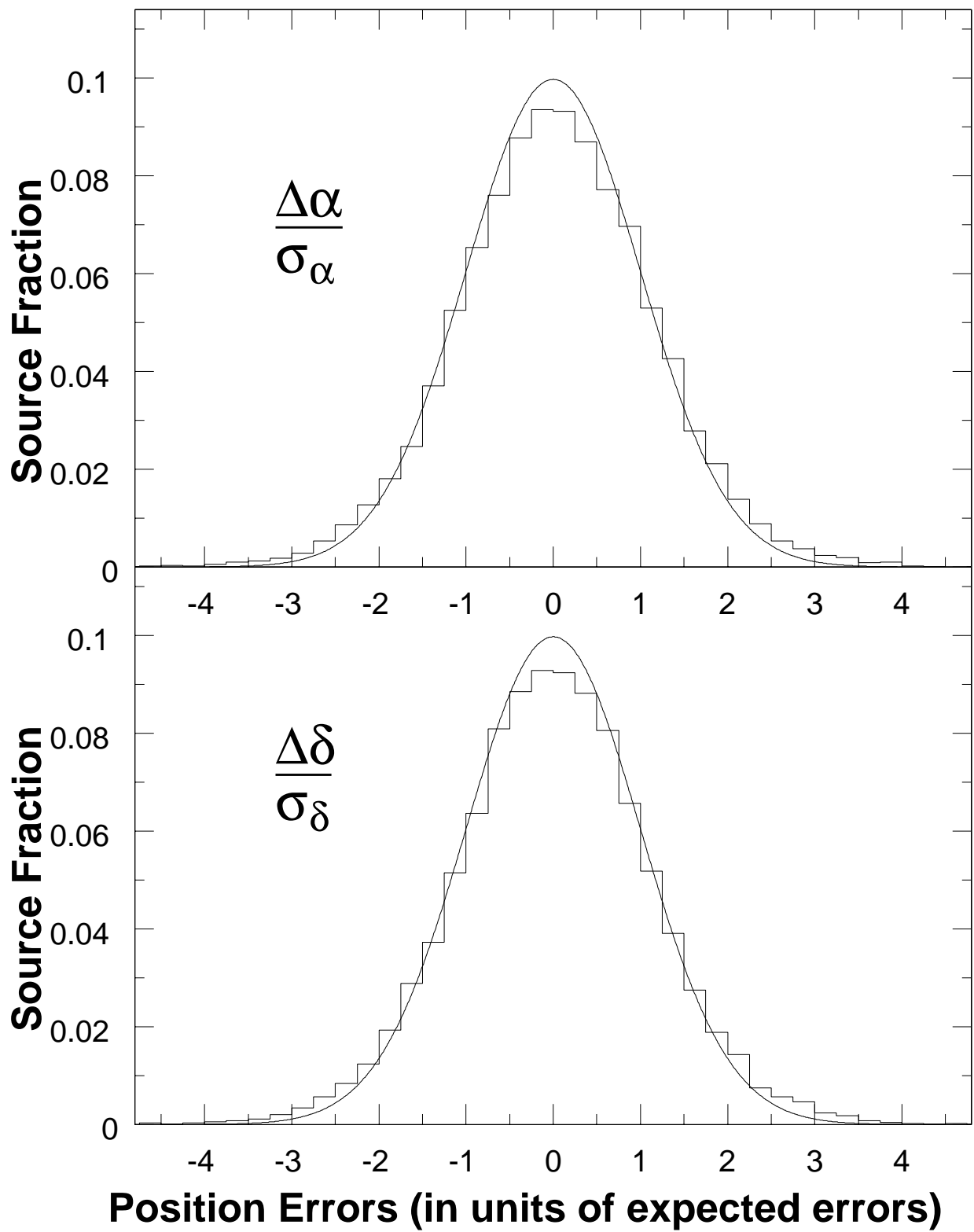}
\caption{
Histograms of the offsets in right ascension ($\Delta\alpha$) and 
declination ($\Delta\delta$) in units of the predicted position errors 
for each source.  Only weak sources with peak flux less than 
10$\sigma$ were included.  The curves plotted over the histograms
are normalized Gaussians with an RMS equal to 1, which is the theoretically
expected distribution.
\label{dpos.weak.fig}}
\end{figure}

To check the validity of these error estimates, we again compared with the 
NVSS positions, but we used relatively weak sources so that the 
source-fitting errors are significant.  Sources were selected with the 
same criteria used in Section \ref{cal.pos.errors}, only this time we only 
selected sources with $5\sigma < I_p < 10\sigma$.  There 
were 18,490 sources that met these criteria, and the resulting position 
error estimates are shown in Figure \ref{dpos.weak.fig}.  All source offsets
($\Delta\alpha$, $\Delta\delta$) were divided by the calculated position 
errors for that source ($\sigma_\alpha$, $\sigma_\delta$).  The ratios of 
actual source offsets to calculated errors, $\Delta\alpha/\sigma_\alpha$ 
and $\Delta\delta/\sigma_\delta$, both had RMS values of 1.12.  These 
values are close to the expected values of unity, and part of the reason 
they are somewhat
higher is likely due to actual source centroid shifts between 74~MHz and 
1.4~GHz.  Therefore we conclude that our error estimates are reasonable.  
In our catalog, we multiplied our position error estimates by 1.12.

\subsection{Source Sizes}

Deconvolved FWHM source sizes, $\phi_M$ and $\phi_m$, can be calculated 
as follows:
\begin{mathletters}
\begin{eqnarray}
\label{dmaj}
\phi_M = \sqrt{(\theta_M^2 - \theta_b^2)} \\
\label{dmin}
\phi_m = \sqrt{(\theta_m^2 - \theta_b^2)} 
\end{eqnarray}
\end{mathletters}
where $\theta_M$ and $\theta_m$ are the fitted FWHM source sizes and 
$\theta_b = 80''$ is the FWHM of the circular restoring beam.  Constraining
the fitted sizes to be larger than the restoring beam guarantees that the
deconvolved source sizes are zero or greater.  The position angle, 
$\phi_{PA}$, is simply the original fitted value.  Equations \ref{dmaj} and 
\ref{dmin} do not take into account imaging artifacts that increase the 
fitted size of even a point source beyond that of the restoring beam, such 
as time-average smearing, bandwidth smearing or, as will be discussed in 
Section~\ref{smear.sec}, ionospheric smearing.  Therefore, values of 
$\phi_M$ and $\phi_m$ are only valid when they are greater than any of these
effects are likely to produce.  As we estimate in Section~\ref{smear.sec} 
this value is about $45''$, which is larger than at least 95\% of ionospheric
smearing effects and also larger than any bandwidth or time-average smearing.
Below $45''$, $\phi_M$ and $\phi_m$ should be treated as upper limits on 
the deconvolved source size.

The errors in the fitted sizes are Gaussian distributed, with RMS error 
given by:
\begin{mathletters}
\begin{eqnarray}
\label{thetaM.error.eqn}
\sigma^2(\theta_M) = \frac{2\theta_M^2}{\rho^2} + \epsilon_o^2\theta_b^2 \\
\label{thetam.error.eqn}
\sigma^2(\theta_m) = \frac{2\theta_m^2}{\rho^2} + \epsilon_o^2\theta_b^2
\end{eqnarray}
\end{mathletters}
where the exponents used for calculating $\rho$  
from Equation \ref{SNR} are $\alpha_M = 5/2$ and $\alpha_m = 1/2$ for 
calculating $\sigma^2(\theta_M)$, and $\alpha_M = 1/2$ and $\alpha_m = 5/2$ 
for calculating $\sigma^2(\theta_m)$ \citep{condon97}.  The term with 
$\epsilon_o$ is a calibration uncertainty which we will estimate in 
Section~\ref{smear.sec}.  

We consider a source to be significantly resolved if the fitted $\theta_M$ 
is larger than the beam size by more than 2.33$\sigma(\theta_M)$.  That is 
the threshold for which a true point source would appear resolved only 2\% 
of the time.  If an axis is significantly resolved, we estimate the error of 
the deconvolved value by varying the fitted size by $\pm1$ standard deviation
and taking half of the difference of the resulting upper and lower 
deconvolved values.  This gives:
\begin{mathletters}
\begin{eqnarray}
\sigma(\phi_M) = \frac{1}{2} \left[ \sqrt{(\theta_M + \sigma(\theta_M))^2 - 
\theta_b^2} - \sqrt{(\theta_M - \sigma(\theta_M))^2 - \theta_b^2} \right] \\
\sigma(\phi_m) = \frac{1}{2} \left[ \sqrt{(\theta_m + \sigma(\theta_m))^2 - 
\theta_b^2} - \sqrt{(\theta_m - \sigma(\theta_m))^2 - \theta_b^2} \right].
\end{eqnarray}
\end{mathletters}

If an axis is not significantly resolved, we only report the 98\% 
confidence upper limits to the deconvolved source size along that axis.  
If at least one axis is significantly resolved, we give the position angle,
$\phi_{PA}$ and its error given by:
\begin{equation}
\sigma^2_{PA} = \frac{4}{\rho^2} \left( \frac{\theta_M\theta_m}
{\theta_M^2 - \theta_m^2} \right)^2 
\end{equation}
where the exponents used for calculating $\rho$ from Equation \ref{SNR} 
are $\alpha_M = 1/2$ and $\alpha_m = 5/2$ \citep{condon97}.

\subsubsection{Effects of Ionospheric Smearing on Source Sizes}
\label{smear.sec}

Ionospheric calibration errors cause artificial smearing of sources and 
increase their 
apparent sizes.  To examine this effect we considered VLSS sources that
are detected at high signal-to-noise ratios because fitting errors on 
strong sources are much smaller than ionospheric smearing effects 
typically are.  

The test we conducted was to measure the Strehl ratios of the VLSS 
sources with the highest peak brightnesses.  The Strehl ratio is the ratio 
of the solid angle of the diffraction limited point source response, which 
in our case is simply the restoring beam (with $80''$ FWHM throughout the 
survey) to the actual solid angle of a point source in an image.  It is 
also the amount by which the peak brightness will be degraded in an image.  
We estimated this by comparing the fitted VLSS source sizes to the expected 
source sizes based on convolving the NVSS counterpart source size to VLSS 
resolution.  This test measures not only ionospheric smearing, but also 
time-average smearing and bandwidth smearing.  However, for most sources, 
incorrect ionospheric calibration is the largest potential cause of smearing.

\begin{figure}
\plotone{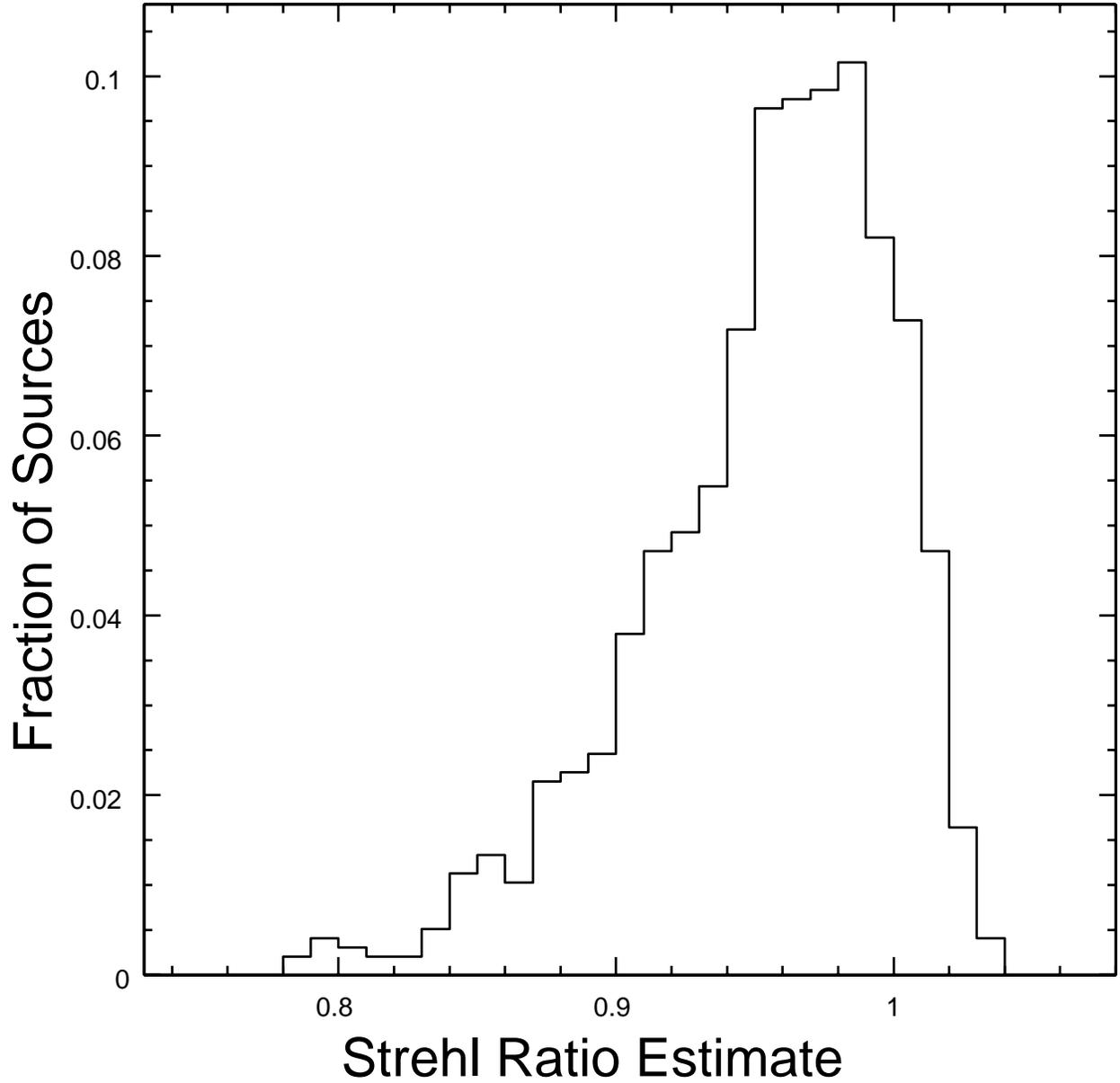}
\caption{
Distribution of estimated Strehl ratios for 975 bright, compact VLSS sources.
The median Strehl ratio is 0.96.  There is a long tail
of values well below the median, which could be due to observations with 
particularly high ionospheric calibration residuals.  Equally plausible is
that some fraction of sources have sizes that really do appear larger at 
74~MHz than at 1.4~GHz because of diffuse steep spectrum emission.
\label{strehl.fig}}
\end{figure}

To avoid other errors, we restricted our sample set to VLSS 
sources with detection levels of at least $60\sigma$ and with NVSS 
counterparts that 
were small enough to imply a VLSS source size of no more than 1.05 times
the restoring beam solid angle.  This resulted in 975 sources, and the 
resulting distribution of Strehl ratios is shown in Figure \ref{strehl.fig}.
Because of measurement errors a small fraction of sources have a Strehl 
ratio above 1, which is unphysical.  Most sources however had
Strehl ratios below 1, with a median value of 0.96.  There is a long tail
of values well below the median, which could be due to observations with 
particularly high ionospheric calibration errors.  Equally plausible is
that some fraction of sources have sizes that really do appear larger at 
74~MHz than at 1.4~GHz because of diffuse steep-spectrum emission.  This 
effect has been previously observed in sources such as Hydra A 
\citep{lane04}.  Finally, one may question whether the use of the sources 
with the highest peak brightnesses, 
many of which were used as calibrator sources, would be biased toward higher
Strehl ratios.  This was checked by taking a similar sample of sources with
peak brightness below the 3.5~Jy/beam cutoff for calibrator selection.  The
weaker source sample actually had a slightly higher median Strehl ratio of 
0.97, but with a much larger spread which is likely caused by the increased 
effects of map noise on weaker sources.  However, it does not seem that the 
source sample is biased toward higher-than-normal Strehl ratios, and therefore
our initial test appears reliable.  Based on the median Strehl ratio of 0.96, 
we use $\epsilon_o = 0.02$ in Equations~\ref{thetaM.error.eqn} and 
\ref{thetam.error.eqn} assuming that on average each axis will have half 
of the residual ionospheric smearing.

This test indicates that ionospheric smearing reduced the peak brightness 
of the typical source by 
about 4\%.  Flux density is not lost, just spread out, so the 
integrated flux density should theoretically not be affected.  However, the
reduction in peak brightness will cause some sources to fall below the 
$5\sigma$ detection threshold, which affects the completeness of the source 
catalog.  

The biggest effect of ionospheric smearing is to increase the 
estimated source sizes.  While a Strehl ratio of 0.96 has a rather small 
effect on the fitted source size, it has a larger effect on the deconvolved
source size estimate.  For a point source, if we take the worst case scenario,
in which the entire source smearing was along one axis, a Strehl ratio of
0.96 would correspond to a deconvolved major axis of $23''$.  A Strehl ratio
of 0.87 (the value of the lowest 95th percentile in the distribution of 
Figure \ref{strehl.fig}) would increase the deconvolved major axis to $45''$.

In the catalog, we report the deconvolved source sizes based only on the 
source fitting without taking into account the effects of ionospheric
smearing.  This is for two main reasons.  First, though we can quantify this
effect statistically for a large number of sources, it is nearly impossible to
estimate the effect for a single source.  Second, for all sources except those
with very high peak brightnesses, any smearing is likely to be small compared 
to the normal 
map noise induced fitting errors, and would not significantly effect the 
reported source sizes.  However, the user of the catalog should be warned that
source sizes smaller than about $45''$ should be treated as upper limits.

\subsection{Flux Densities}

In this section we describe how the source flux densities and their errors
were determined.  The flux-density 
error estimates are quadratic sums of map-noise-induced fitting errors 
(Section~\ref{flux.noise}) and 
intensity-proportional flux errors produced by errors in both the 
flux-density scale and primary-beam corrections 
(Sections \ref{flux.scale} and \ref{PB.error}).  We also correct the flux 
densities for a clean bias which we discuss in Section~\ref{cleanbias}.
In Section~\ref{compare.sec}, we compare our flux values to those from other 
notable low-frequency surveys as a general check of our results.

\subsubsection{Flux-Density Errors from Map Noise}
\label{flux.noise}

The map noise error depends on the effective 
signal to noise ratio, $\rho$, which for flux densities should be calculated 
using Equation \ref{SNR} with $\alpha_M = \alpha_m = 3/2$ \citep{condon97}.
Also, map noise error depends on the number of
degrees of freedom allowed in the Gaussian fit.  Therefore, the best estimate 
for the integrated flux density of a source depends on whether or not the 
source is significantly resolved \citep{condon97}.  There are three cases to 
consider:

1. If both the major and minor axes are significantly resolved, the integrated
flux density is
\begin{equation}
S_i = I_p \left( \frac{\theta_M\theta_m}{\theta_b^2} \right)
\end{equation}
and the RMS uncertainty in $S_i$ is
\begin{equation}
\label{case1}
\sigma(S_i) = S_i 
\left(
\epsilon_S^2 + \frac{2I_p^2}{\rho^2} + 
\frac{\theta_b^2}{\theta_M\theta_m} 
\left[ \frac{\sigma^2(\theta_M)}{\theta_M^2} 
+ \frac{\sigma^2(\theta_m)}{\theta_m^2} \right]
\right)^{1/2}
\end{equation}

2. If the major axis is significantly resolved but the minor axis is not, 
the best estimate for the integrated flux density is
\begin{equation}
S_i = I_p \left( \frac{\theta_M}{\theta_b} \right)
\left( \frac{\theta_m}{\theta_b} \right)^{1/2}
\end{equation}
and the RMS uncertainty in $S_i$ is
\begin{equation}
\label{case2}
\sigma(S_i) = S_i 
\left( 
\epsilon_S^2 + \frac{3I_p^2}{2\rho^2} + 
\frac{\theta_b}{\theta_M} 
\frac{\sigma^2(\theta_M)}{\theta_M^2} 
\right)^{1/2}
\end{equation}

3. If neither the major nor minor axis is significantly resolved, the best 
estimate for the integrated flux density is 
\begin{equation}
S_i = I_p \left( \frac{\theta_M\theta_m}{\theta_b^2}\right)^{1/2}
\end{equation}
and the RMS uncertainty in $S_i$ is
\begin{equation}
\label{case3}
\sigma(S_i) = S_i 
\left( \epsilon_S^2 + \frac{I_p^2}{\rho^2}\right)^{1/2}.
\end{equation}

\subsubsection{Accuracy of the Flux-Density Scale}
\label{flux.scale}

The flux-density scale for each VLSS observation was determined by comparing
scans on Cygnus~A to a pre-existing model that was scaled to the flux density 
determined by \citet{baars77}.
As the 74~MHz VLA is a fairly new system, the accuracy of this method is 
not well determined.  Ideally we could compare our results for very strong 
sources to some pre-existing database of accurate 74~MHz flux densities.  
However, no database of 74~MHz flux densities exists, nor does any database
of low-frequency source spectra with enough accuracy for us to measure the 
accuracy of our own flux-density scale.

Fortunately, Cygnus~A was not the only calibrator we observed.  At times when 
Cygnus~A was not at high enough elevation, two other calibrators, Virgo~A and 
3C~123, were observed in case the calibration from other times on Cygnus~A was
not sufficient.  This never was the case, so these other calibrators were 
never actually used in data reduction.  However, as they are sources
observed on many different days, their derived flux densities on each day
can be used as an internal check of the flux calibration consistency.  
3C~123 was observed on 24 separate days, and the flux-density measurements 
give a mean value of $390.5$~Jy with an RMS scatter of $15.7$~Jy, 
indicating that the flux-density scale
is consistent to within 4\%.  This 
measurement agrees well with the value predicted by \citet{kuhr81} which 
is 387.6~Jy indicating that flux-density scale remains linear even at 
Cygnus~A flux density levels.  Virgo~A was observed on 42 days, yielding a 
mean flux density of $1911.5$~Jy with an RMS scatter of $119.2$~Jy
which indicates that the flux-density scale 
is consistent to within 6\%.  This is somewhat different than the predicted 
\citet{kuhr81} value of 2281.4~Jy.  However, the \citet{kuhr81} spectrum 
is even farther off from its own 80~MHz data point of 1519~Jy, indicating 
that their spectrum may not be a good fit to the data.  The fact that Virgo~A 
is a large, heavily resolved source may also be a factor.  To be on the 
conservative side, and also because Virgo~A was observed more often, we 
consider the 6\% to be a reasonable estimate of the accuracy of our 
flux-density scale.

\subsubsection{Total Intensity-Proportional Flux-Density Errors}
\label{PB.error}

The errors in the flux-density scale, discussed in Section \ref{flux.scale},
are only part of the total intensity-proportional flux errors.  The other 
significant component is the primary-beam correction error.  The primary 
beam of a VLA antenna at 74~MHz is not perfectly understood and is probably 
not symmetric.  The analysis of Section \ref{flux.scale} only considered
sources at the center of the field of view, for which primary-beam corrections
were not necessary.  In this section we analyze the contribution of 
primary-beam correction errors to the intensity-proportional flux error of
any VLSS sources.

To determine the total intensity-proportional flux errors for a general VLSS
source, we take advantage of the fact that most sources have been observed
in two or three different pointings because of the overlapping nature of the
observational grid.  Thus the original field maps, prior to their combination 
with adjacent fields in the mosaic, contain multiple flux-density measurements 
with independent flux-density calibrations and primary-beam 
corrections.  For sources detected at sufficiently high signal-to-noise 
ratios, map noise errors are negligible compared with intensity-proportional
flux errors.  Therefore, for strong sources, the differences among 
flux-density measurements made in overlapping fields are due almost
entirely to intensity-proportional flux errors.

We considered all sources detected with signal-to-noise ratios of at least 
60 and that had fitted Gaussian major axes less than 120$''$ (1.5 times 
the restoring beam) to remove sources that are large enough that they may 
not be well fit by a Gaussian.  There were 1,126 such sources and 1,787 pairs
of flux-density measurements of the same source from different fields.  
The root-mean-square of the fractional differences in these measurement pairs 
indicates a total intensity-proportional flux error of 12.1\%.  This includes 
both the
flux-density scale errors and primary-beam correction errors.  Taking the 
estimated flux-density scale error of 6\% from Section~\ref{flux.scale}, the
primary-beam correction error alone is about 10.5\%.

The 12.1\% value for the intensity-proportional flux error is almost 
certainly an overestimate for two reasons.  First,
because we only consider sources observed in two or more adjacent fields,
we are biased toward sources near the outer region of the primary beam, 
for which primary-beam corrections will have greater than average errors.  
Second, this is the estimated error for a single observation.  Most VLSS
sources are a weighted sum of adjacent field images, and so the averaged
error will be somewhat less than the error in a single field.  In fact, 
the comparisons to other catalogs presented later in Section~\ref{compare.sec} 
do indicate that the intensity-proportional flux errors are probably somewhat 
smaller than this.  In particular, the comparisons to 6C and 8C sources 
shows that our intensity-proportional error can be no higher than 10.4\% 
and even this includes the errors from both of those surveys as well as 
errors in interpolation which assume a constant spectral index.  Therefore,
we use 10\% as a conservative estimate of the intensity proportional error
and set $\epsilon_S$ = 0.10 for the flux-density 
error determinations of Equations \ref{case1}, \ref{case2} and \ref{case3}.

It should be noted that $\epsilon_S$ = 0.10 is the measure of the 
consistency of our flux-scale and primary-beam corrections with respect to 
the \citet{baars77} 
scale.  That scale itself has an absolute accuracy that they estimate to be 
5\% at frequencies below about 300~MHz.

\subsubsection{Clean Bias}
\label{cleanbias}

Images deconvolved using the CLEAN algorithm are known to suffer from a 
``clean bias''.  This occurs because, as cleaning proceeds to deeper levels, 
the probability increases that a sidelobe of a source or a noise fluctuation 
(or a combination of these) can produce a peak higher than any remaining flux
in the image.  Cleaning this false source results in flux being subtracted 
from the true source that produced that sidelobe.  Therefore
the clean bias results in the flux densities of sources being systematically 
reduced.  The magnitude of the bias is independent of the flux density of 
sources, but is known to increase with higher map noise.  

We have conducted a simulation to determine the clean bias in our images.  
This was done by adding artificial point sources to the $uv$-data and 
comparing the input flux densities to the measured flux densities in the 
resulting maps.  This is a standard way to measure the clean bias, however
the fact that we conducted field-based ionospheric calibration during imaging
complicated this simulation.  Because ionospheric phase corrections are 
applied to the data during imaging, we had to apply the inverse of these 
corrections to the artificial sources before adding them to the $uv$-data 
in order to produce the intended artificial point sources in the final maps.

In all, we produced 2362 artificial sources at random positions in fields 
chosen from the survey data to represent a wide range of right ascension and 
declination.  The flux densities varied randomly 
from 1 to 10 Jy.  The final results showed a systematic bias between the 
input flux densities and the map results, confirming the clean bias.  The 
measured clean bias was independent of the input flux density, but did 
increase with local RMS maps noise, $\sigma$.  Fitting a model of the 
clean bias where its value is proportional to $\sigma$, we found the average
clean bias reduces the flux densities of point sources by 1.39$\sigma$.  
We have corrected the source catalog for this bias by adding this amount, 
to the peak brightness, $I_p$, of each source when calculating the flux 
density, $S_i$, as described in Section~\ref{flux.noise}.  Though larger 
than the local RMS map 
noise, the clean bias is still smaller than the flux density error for any
source because of the intensity-proportional errors.  Therefore, even with 
the clean bias correction, flux density errors are still dominated by the 
map noise errors and intensity-proportional errors calculated previously.

The magnitude of the clean bias we find is similar to that of other surveys.
It is a bit higher than the value found in the NVSS survey, which is about 
0.67 times the typical map noise \citep{condon98}.  It is a bit less than 
the value found in the FIRST survey, which is about 1.67 times the typical 
map noise \citep{becker95}.  We note that both the NVSS and FIRST use a 
constant clean bias correction, rather than one that is proportional to the 
local RMS noise as we do, and the ratios mentioned are relative to the 
average noise levels of those surveys.

\subsubsection{Comparisons with other low-frequency data}
\label{compare.sec}

Though it proved impossible to use other low-frequency catalogs to predict 
74~MHz flux densities of enough sources with enough precision to test the 
accuracy of our own flux-density scale, comparisons to other low-frequency 
data are still of interest.  In this section, we describe a few of the 
comparisons we conducted.

\begin{figure}
\plotone{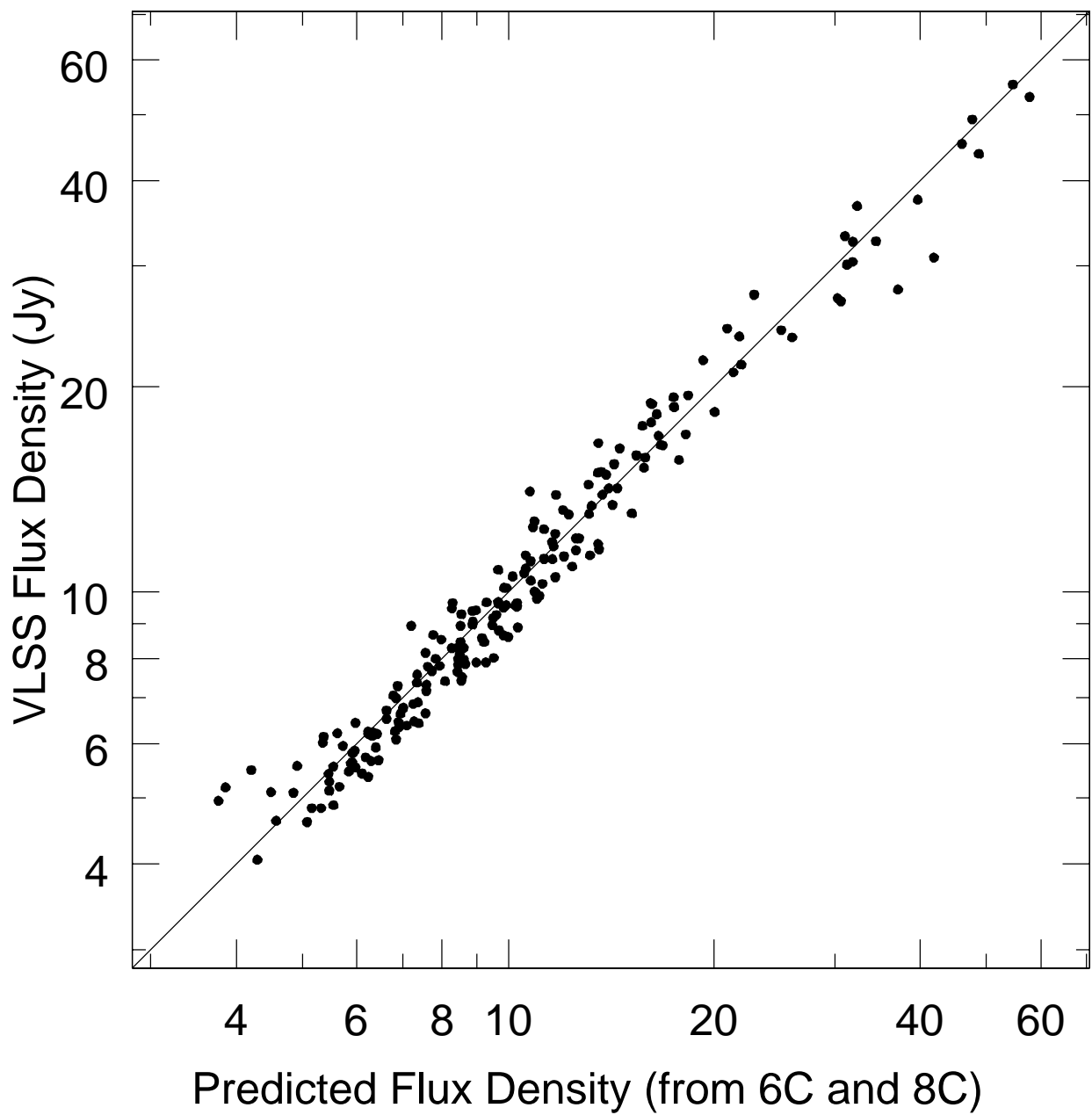}
\caption{
Comparisons of predicted flux densities versus measured flux densities for 
bright VLSS sources.  Predicted flux densities at 73.8 MHz were calculated 
by interpolating between the 6C and 8C flux densities for the 201 bright 
VLSS sources that had counterparts in both the 6C and 8C surveys.  The 
average ratio of the VLSS flux densities to the predicted values was 
$0.99\pm0.01$ with a scatter of $\pm10.4$\%.  The straight line represents 
locations for which the predicted and measured flux densities are equal.
\label{flux.compare.fig}}
\end{figure}

\begin{description}

\item[\citet{kuhr81} Spectra:] 
We have compared the flux densities of strong VLSS sources to 
their expected
values based on fitted spectra provided in \citet{kuhr81}, which also uses
the flux-density scale of \citet{baars77}.  We used sources
with VLSS integrated flux densities between 15 Jy $\leq S_{74} \leq$ 200 Jy 
that also had \citet{kuhr81} spectra based on data including at least one 
measurement below 100~MHz.  There were 94 such sources, and the ratio 
of VLSS flux densities to the \citet{kuhr81} predicted flux densities had a 
RMS scatter of $\pm20$\% and a mean value of $0.87 \pm 0.02$. 

This ratio seems rather low, and the scatter quite high.  One problem with 
this method is that the errors in most \citet{kuhr81} spectra
tend to be quite large at the low-frequency end.  In particular, the fitted
spectra were often quite far off the actual measurement values below 100~MHz.
In fact, for sources with 80~MHz measurements, the difference between the 
\citet{kuhr81} fitted spectrum and the 80~MHz value was often greater than 
the difference between the 74~MHz VLSS value and the fitted spectrum.

\item[80~MHz data:]
Because of the problems experienced in comparing VLSS flux densities to the 
\citet{kuhr81} spectra, we chose to simply compare the VLSS measurements 
directly to the 80~MHz measurements.  The 80~MHz measurements were taken 
with the Culgoora radioheliograph \citep{slee73,slee75,slee77} and adjusted
to the \citet{baars77} flux-density scale by \citet{kuhr81}.  We adjusted 
these 80~MHz values to predicted 74~MHz values by assuming a spectral index 
of $\alpha = -0.8$.  For the 70 sources with 80~MHz 
measurements in the \citet{kuhr81} spectra, the ratio of VLSS to the 
resulting predicted flux densities had an RMS scatter of $\pm12$\% and a mean 
value of $0.93 \pm 0.01$, which is a better ratio and scatter than 
the comparisons to the \citet{kuhr81} spectra.  Because this value depends
on the assumed average spectral index, it should be considered as a lower 
limit, as a significant fraction of sources could have flattened due to 
synchrotron self-absorption and would have spectra flatter than the typical 
value of $\alpha = -0.8$.  If so, this would raise mean flux-density ratio.  
For example, if we assume an average spectral index of $\alpha = -0.4$, the 
derived mean flux-density ratio becomes $0.96 \pm 0.01$.

\item[8C to 6C interpolation:]
Another comparison to external data was done by predicting the flux at 74~MHz
by interpolating between values from the 38~MHz 8C catalog 
\citep{rees90} and the 151~MHz 6C catalog \citep{hales88}.  Choosing VLSS 
sources that were detected above at least the 60$\sigma$ level 
and also had single counterparts within 120$''$ in both the
8C and 6C catalogs resulted in a sample of 201 sources.  The 8C and 6C 
catalogs are based on the flux-density scale of \citet{roger73}, and so 
we adjusted the interpolated flux-density value to the \cite{baars77} scale
according to the flux ratios of Cygnus~A.  Figure~\ref{flux.compare.fig} 
shows a comparison of the flux densities of the VLSS 
sources versus the interpolated predictions.  For these 
sources the average ratio of VLSS to expected flux density had an RMS 
scatter of 
$\pm10.4$\% and a mean value of $0.99 \pm 0.01$.  This test had a final ratio 
quite close to unity as well as a low scatter.  After taking into account the 
errors in the 8C and 6C catalogs, this result seems consistent with our 
flux-density scale error estimate of 6\%, and actually lower than our 
admittedly conservative estimate of the total intensity-proportional 
flux-density error.

\end{description}

\section{Completeness of the Catalog}

In this section we discuss the point-source completeness of the VLSS catalog.  
Because the RMS noise levels are not constant throughout the survey region, 
we estimated not the total completeness, but the differential completeness as 
a function of signal-to-noise ratio (the 
ratio of the peak brightness of a point source to the local RMS noise level).  
This was estimated through
simulations in which artificial sources were added at random positions to 
the actual VLSS images.  The same source-finding methods were then applied
to the maps to see what fraction of artificial sources were ``detected''.

\begin{figure}
\plotone{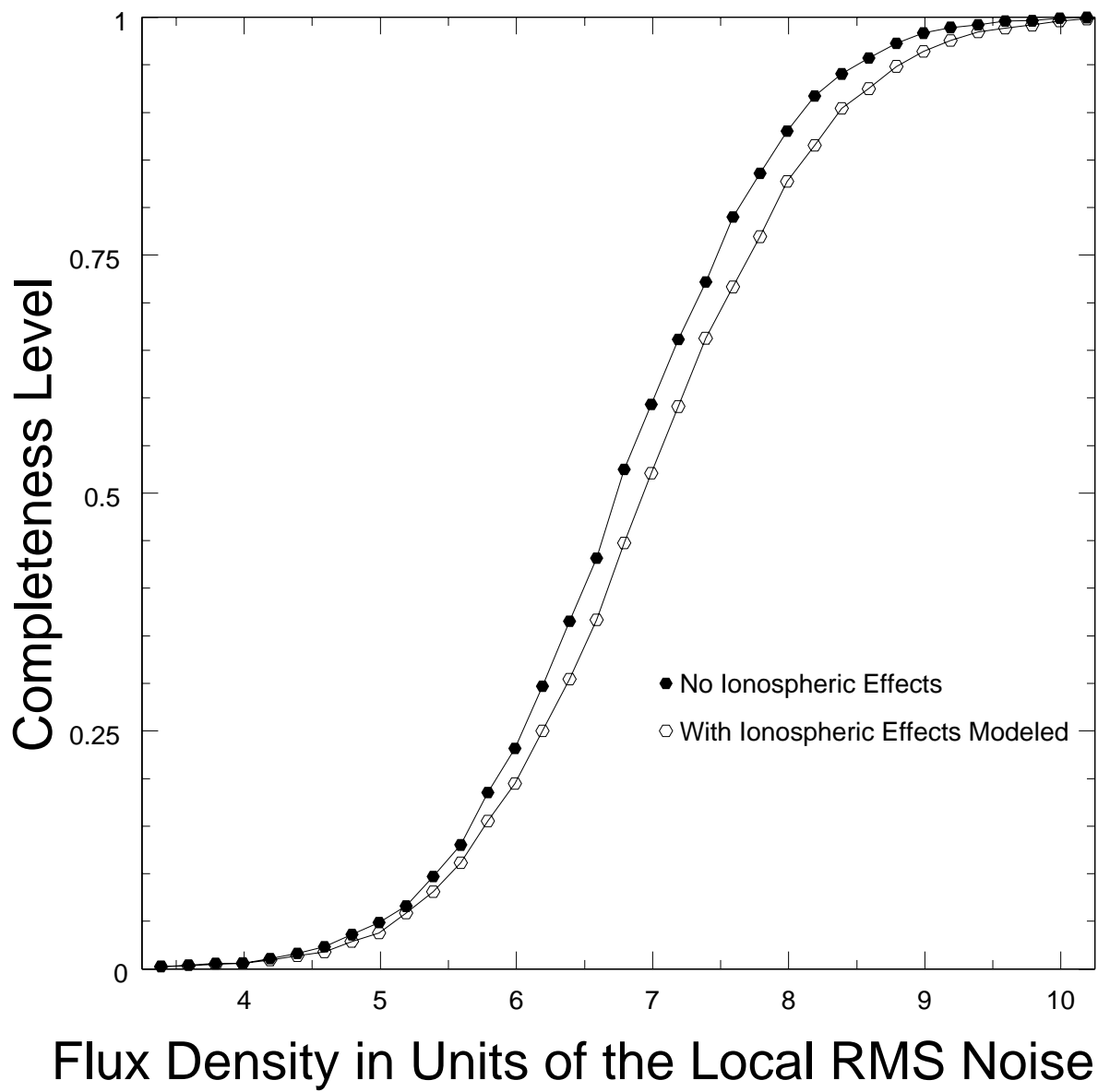}
\caption{
Differential completeness of the VLSS catalog for point sources as a 
function of the ratio of peak 
flux density to local RMS noise level.  (This completeness measure does not
apply to the regions surrounding the strongest sources for which higher source
selection criteria were used.)  The filled points are calculated 
from simulations that do not take ionospheric smearing into account, while the 
open points are from simulations that do account for this.  The entire curve 
is shifted to the right by the clean bias which is 1.39 in units of the 
local RMS noise.
\label{complete.fig}}
\end{figure}

Simulations were first done for point sources without taking into account 
any ionospheric source smearing.  Gaussians were added to the 
VLSS images with the same dimensions as the restoring beam (circular with 
FWHM of 80$''$).  The peak brightnesses of the artificial sources
were set to various multiples of the local RMS noise levels and reduced by 
the clean bias of 1.39$\sigma$ as determined in Section~\ref{cleanbias}.  
For a given signal-to-noise ratio, 32 artificial sources were added at random 
positions to each of the 279 $14^\circ\times14^\circ$ VLSS images that had 
no blanked regions, for a total sample of 8,928 artificial sources.  The 
fraction
of these that were ``detected'' according to our source-finding criteria was
taken to be the differential completeness level at that signal-to-noise 
ratio.  These values are plotted in Figure~\ref{complete.fig} as the filled 
points.  
Because the source finding criteria used included the $5\sigma$ fitted 
peak brightness, it does not apply to the regions surrounding high peak
brightness sources for which a $6\sigma$ selection was used as described
in Section~\ref{falsesources.sec}

We performed a second set of simulations which aimed to factor in residual
ionospheric smearing.  The Strehl ratio distribution of strongly detected, 
compact VLSS sources found in Section~\ref{smear.sec} was used as a guide.  
If one assumes that the majority of the smearing occurs on one axis, that 
Strehl ratio roughly
corresponds to a mean deconvolved major axis of 22.9$''$ with an RMS deviation
of 14.2$''$.  Therefore we repeated the same simulation described above, but
this time convolving each artificial source with a Gaussian with major axis
equal to a Gaussian random variable with a mean of $22.9''$ and RMS of 
$14.2''$, a minor axis equal to zero and a PA set randomly between $0^\circ$ 
and $360^\circ$.  (If the random major axis was less than zero, it was replaced
with zero.)  The peak brightness of each artificial source was thus lowered 
by the same factor that its area was increased due to smearing.  The resulting
differential completeness levels are also shown in Figure~\ref{complete.fig} 
as the open points.  This simulation indicates that ionospheric 
smearing has most likely reduced the completeness of the survey to some 
degree, with the effect being greatest for the faintest sources.  
The completeness level for $6\sigma$ point sources is lowered by 16\%, however
the difference falls to 12\% for $7\sigma$ sources and 6\% for $8\sigma$ 
sources.  A much larger effect on completeness is caused by the clean bias, 
which shifts the entire differential completeness curve of 
Figure~\ref{complete.fig} to the right by 1.39 in units of the local RMS 
noise level.  Taking into account both ionospheric smearing and the clean 
bias, the 50\% point-source detection limit of the survey is roughly 
$7\sigma$, or about 0.7~Jy for a typical noise level of 0.1~Jy/beam.

\section{Accessing the Data}

As the VLSS was conducted as a service to the astronomical community, 
we have publicly released all images and catalogs as soon as they were 
reduced and verified.  Along with an overall description of the VLSS project
and its current status, all verified data can be obtained at the 
VLSS website:\\

{\tt http://lwa.nrl.navy.mil/VLSS}

\noindent
The data are available in several forms.  First, all images in the grid of 
the original large $\approx14^{\circ}\times14^{\circ}$ combined images are 
available to be directly downloaded as FITS files.  Further, we have made 
available an online postage-stamp server in which a smaller image around a 
point of interest can be obtained for any user-entered coordinates within the
currently completed survey region.  

The current VLSS catalog file can be downloaded as well.  Also provided 
is browser software which uses the catalog file to search for sources within 
a given radius of given positions using various criteria and gives the final 
parameters and their error estimates.  A small sample the output of the VLSS 
catalog browser is shown in Table \ref{tab.catalog}. 

\section{Conclusion and Future Work}

We have nearly completed a 74~MHz survey of the entire sky visible from the 
northern hemisphere.  This was done by overcoming the large ionospheric 
effects which have been the main obstacle to wide-field arcminute-resolution 
imaging at this frequency for the last 50 years.  We now have high-quality 
74~MHz images for roughly $3\pi$ sr. of sky, yielding nearly 70,000 
sources detected at the $5\sigma$ level or greater in the images.  All 
VLSS data are available publicly at the VLSS website:\\

{\tt http://lwa.nrl.navy.mil/VLSS}

In terms of both sky coverage and resolution, we now have produced a low 
frequency equivalent to the most comprehensive radio sky survey yet 
produced, the NVSS.  The one aspect that keeps the VLSS from being a full 
equivalent to the NVSS is sensitivity.  The VLSS has a typical RMS noise 
level roughly 220 times higher than the NVSS, and even allowing for an average 
spectral index of $\alpha = -0.8$, it is still less sensitive than the NVSS 
by a factor of 20.   As discussed in Section~\ref{sensitivity.sec}, the 
limiting factor in VLSS sensitivity is sidelobe confusion.
A telescope designed specifically for low frequencies could overcome this 
problem to a large extent with better forward gain, a smaller field of view,
and higher resolution.  Yet, even without sidelobe confusion, thermal noise 
still prevents low-frequency observations from attaining sensitivity 
anywhere near that of cm-wavelengths.  This is because of the very high sky 
noise temperature at low frequencies caused by Galactic synchrotron emission.  
The only solution to this is a telescope 
with vastly more collecting area than the VLA.  A number of low frequency 
instruments are being planned (ie. LWA, LOFAR) which will have huge 
collecting areas.  It is these instruments which will truly bring 
low-frequency observations up to the level of cm-wavelength capabilities.  
It is hoped that the VLSS project will provide both inspiration (in the form 
of observational experience and science) and information (in the form of a 
sky model and ionospheric data) for the design and operation of these new 
instruments.

\section{Acknowledgments}

Basic research in radio astronomy at the Naval Research Laboratory is 
supported by the office of Naval Research.  The National Radio Astronomy 
Observatory is a facility of The
National Science Foundation operated under cooperative agreement by
Associated Universities, Inc.  We thank Joseph Helmboldt for helping to 
diagnose errors in the source flux-density measurements.

\clearpage

\begin{deluxetable}{ll}
\tablecaption{VLSS Project Parameters\label{tab.parms}}
\tablehead{
\colhead{Parameter} & \colhead{Value}
}
\tablewidth{0pt}
\startdata
Frequency & 73.8 MHz \\
Total Bandwidth & 1.56 MHz \\
Channel Width & 12.2 kHz \\
Channel Width (averaged) & 122 kHz \\
Resolution & $80''$ \\
RMS noise level & 100 mJy/beam (typical) \\
Survey Region & all sky above $\delta > -30^{\circ}$ \\
Survey Area & $3\pi$ sr. \\
Sources Detected ($\geq5\sigma$) & 68,311 (as of 2007)\\ 
\enddata
\end{deluxetable}

\clearpage

\begin{deluxetable}{lccr}
\tablecaption{Observation Dates for the VLSS\label{tab:Obsdates}}
\tablehead{
\colhead{Project}  &  \colhead{Dates}    &   
\colhead{Configuration}  &       \colhead{Total Time (hrs)}
}
\startdata
AP 397   &  9,26 Feb 2001 &       B   &        36 \\
AP 441   &  8-20 Jun 2002 &      B    &        64 \\
AP 452   &  20 Sep - 6 Oct 2003  &  BnA  &     69 \\
   ~~    &  18 Oct - 7 Dec 2003   &  B   &    283.5 \\
   ~~    &  24 Jan - 14 Feb 2005  &  BnA  &     96 \\
   ~~    &  19 Feb - 7 Apr 2005  &  B    &    289.5 \\
   ~~    &  3  Jun - 9 Jun 2006   & BnA       &     43.5 \\
  ~~     &   Fall 2007            & BnA       &     10 \\
AP 509    &  17 Jun - 29 Jun 2006 &  B         &     25.5 \\
   ~~       &  Fall 2007            & B         &     10 \\
\enddata
\end{deluxetable}

\clearpage

\begin{deluxetable}{lrrrrrrrrr}
\tablewidth{0pt}
\tabletypesize{\tiny}
\tablecaption{Sample Section from the VLSS Source Catalog\label{tab.catalog}}
\tablehead{
\colhead{Source} & 
\colhead{$\alpha$ (J2000)} & 
\colhead{$\delta$ (J2000)} & 
\colhead{Flux} & 
\colhead{} & 
\colhead{Size} & 
\colhead{} & 
\colhead{Field} & 
\colhead{X$_{pix}$} & 
\colhead{Y$_{pix}$} 
\\
\colhead{} & 
\colhead{(h, m, s)} & 
\colhead{($^{\circ}$,$'$,$''$)} & 
\colhead{Jy} & 
\colhead{maj. $('')$} & 
\colhead{min. $('')$} & 
\colhead{PA $(^{\circ})$} & 
\colhead{} & 
\colhead{} & 
\colhead{} 
}
\startdata
VLSS J0917.9$+$5853 & 09 17 59.45  &  $+$58 53 26.3 &       1.28 & $<$57.1 & $<$56.8 &  & 0900$+$550 &     691 &    1596\\
 &        0.59 &            4.7 &       0.15 &  &  &  & & & \\
VLSS J0917.9$+$0034 & 09 17 59.51  &  $+$00 34 48.3 &       1.13 & $<$62.8 & $<$62.6 &  & 0920$+$050 &    1097 &     389\\
 &        0.35 &            5.3 &       0.14 &  &  &  & & & \\
VLSS J0918.0$+$3001 & 09 18 01.15  &  $+$30 01 28.5 &       0.94 & $<$61.2 & $<$60.4 &  & 0936$+$350 &    1585 &     322\\
 &        0.38 &            5.1 &       0.12 &  &  &  & & & \\
VLSS J0918.0$-$0303 & 09 18 01.38  &  $-$03 03 53.1 &       0.69 & $<$86.1 & $<$81.8 &  & 0920$-$050 &    1096 &    1304\\
 &        0.52 &            7.6 &       0.11 &  &  &  & & & \\
VLSS J0918.0$-$1303 & 09 18 02.53  &  $-$13 03 13.8 &       2.41 & $<$82.6 & $<$53.2 &  & 0920$-$150 &    1094 &    1305\\
 &        0.31 &            4.8 &       0.28 &  &  &  & & & \\
VLSS J0918.0$-$1205 & 09 18 03.63  &  $-$12 05 02.0 &     579.59 & 186.0 & 78.1 & 9.1 & 0920$-$150 &    1093 &    1445\\
 &        0.20 &            3.0 &      58.05 &       1.8 &       2.3 &      0.10 & & & \\
VLSS J0918.0$+$4311 & 09 18 04.32  &  $+$43 11 57.2 &       1.12 & $<$62.0 & $<$61.8 &  & 0936$+$450 &    1495 &     779\\
 &        0.47 &            5.2 &       0.14 &  &  &  & & & \\
VLSS J0918.0$-$2402 & 09 18 04.98  &  $-$24 02 30.8 &       2.95 & $<$50.4 & $<$38.6 &  & 0900$-$250 &     431 &    1153\\
 &        0.25 &            3.4 &       0.31 &  &  &  & & & \\
VLSS J0918.1$+$6537 & 09 18 07.02  &  $+$65 37 16.8 &       1.39 & $<$51.5 & $<$51.5 &  & 0920$+$650 &    1053 &    1115\\
 &        0.67 &            4.3 &       0.16 &  &  &  & & & \\
VLSS J0918.1$+$3225 & 09 18 07.17  &  $+$32 25 28.1 &       1.49 & $<$79.2 & $<$50.2 &  & 0936$+$350 &    1568 &     666\\
 &        0.35 &            4.5 &       0.17 &  &  &  & & & \\
VLSS J0918.1$+$1227 & 09 18 09.85  &  $+$12 27 46.1 &       6.08 & 31.8 & $<$38.2 & 84.7 & 0920$+$150 &    1090 &     660\\
 &        0.20 &            3.1 &       0.64 &       6.4 &  &     24.34 & & & \\
VLSS J0918.1$-$0841 & 09 18 10.84  &  $-$08 41 34.9 &       2.19 & 174.8 & $<$116.1 & 41.5 & 0920$-$150 &    1090 &    1931\\
 &        1.06 &           17.2 &       0.55 &      50.2 &  &      9.97 & & & \\
VLSS J0918.1$+$2950 & 09 18 11.41  &  $+$29 50 25.2 &       1.39 & $<$91.0 & $<$49.9 &  & 0900$+$250 &     458 &    1731\\
 &        0.31 &            5.1 &       0.16 &  &  &  & & & \\
VLSS J0918.1$-$3010 & 09 18 11.50  &  $-$30 10 17.5 &       1.72 & $<$83.4 & $<$73.8 &  & 0936$-$350 &    1579 &    1707\\
 &        0.41 &            5.2 &       0.21 &  &  &  & & & \\
VLSS J0918.2$+$6451 & 09 18 13.55  &  $+$64 51 07.2 &       3.61 & 68.0 & $<$34.6 & 31.2 & 0920$+$650 &    1052 &    1004\\
 &        0.51 &            3.6 &       0.40 &       7.2 &  &      4.83 & & & \\
VLSS J0918.2$+$8111 & 09 18 13.88  &  $+$81 11 27.8 &       1.40 & $<$71.9 & $<$58.4 &  & 0800$+$850 &     602 &     550\\
 &        2.10 &            5.2 &       0.17 &  &  &  & & & \\
VLSS J0918.2$-$0053 & 09 18 14.01  &  $-$00 53 46.6 &       0.86 & $<$139.5 & $<$76.2 &  & 0920$-$050 &    1089 &    1615\\
 &        0.62 &            9.2 &       0.14 &  &  &  & & & \\
VLSS J0918.2$-$3048 & 09 18 16.11  &  $-$30 48 50.7 &       4.01 & $<$51.1 & $<$36.3 &  & 0936$-$350 &    1573 &    1615\\
 &        0.25 &            3.4 &       0.42 &  &  &  & & & \\
VLSS J0918.2$+$6844 & 09 18 17.01  &  $+$68 44 06.9 &       1.01 & $<$56.1 & $<$55.6 &  & 0920$+$650 &    1047 &    1563\\
 &        0.83 &            4.6 &       0.12 &  &  &  & & & \\
VLSS J0918.3$-$0005 & 09 18 23.32  &  $-$00 05 10.1 &       0.96 & $<$135.1 & $<$73.0 &  & 0920$-$050 &    1083 &    1732\\
 &        0.67 &            6.7 &       0.15 &  &  &  & & & \\
VLSS J0918.4$-$2709 & 09 18 24.02  &  $-$27 09 09.7 &       5.15 & 33.2 & $<$40.0 & 125.9 & 0900$-$250 &     436 &     705\\
 &        0.23 &            3.2 &       0.55 &       7.4 &  &     26.12 & & & \\
VLSS J0918.4$-$3133 & 09 18 24.17  &  $-$31 33 32.7 &       3.36 & 56.3 & $<$46.1 & 86.5 & 0936$-$350 &    1564 &    1508\\
 &        0.32 &            3.7 &       0.43 &      11.4 &  &     11.90 & & & \\
VLSS J0918.4$+$7143 & 09 18 27.66  &  $+$71 43 56.6 &       0.56 & $<$99.5 & $<$81.8 &  & 1000$+$750 &    1491 &     596\\
 &        1.82 &            7.7 &       0.09 &  &  &  & & & \\
\enddata
\tablecomments{
Description of columns: \\ 
1. Source Name (in the form ``VLSS JHHMM.m+DDMM'' as per International Astronomical Union (IAU) recommendations) \\ 
2. Right Ascension (J2000) \\ 
3. Declination (J2000) \\ 
4. Integrated Flux Density in Jy \\ 
5. Size of Deconvolved Major Axis in arcseconds, or the $2\sigma$ upper limit if the source is unresolved \\ 
6. Size of Deconvolved Minor Axis in arcseconds, or the $2\sigma$ upper limit if the source is unresolved \\ 
7. PA of Major axis in degrees. \\ 
8. VLSS Image Field \\ 
9. X pixel of source in VLSS Image Field \\ 
10. Y pixel of source in VLSS Image Field \\ 
}
\tablecomments{
Errors, if applicable, are given in the same units in the next row. \\}
\end{deluxetable}


\clearpage

\clearpage

\clearpage

\clearpage

\clearpage

\clearpage

\clearpage

\clearpage

\clearpage

\clearpage

\clearpage

\clearpage

\clearpage

\clearpage

\clearpage

\end{document}